\documentclass[graybox,natbib,10pt,vecphys]{svmult}
\usepackage{mathptmx}       
\usepackage{amsmath}
\usepackage{helvet}         
\usepackage{courier}        
\usepackage{type1cm}        
\usepackage{makeidx}         
\usepackage{graphicx}        
\usepackage{multicol}        
\usepackage[bottom]{footmisc}

\def\ltsima{$\; \buildrel < \over \sim \;$}    
\def\lesssim{\lower.5ex\hbox{\ltsima}}           
\def\gtsima{$\; \buildrel > \over \sim \;$}    
\def\gtrsim{\lower.5ex\hbox{\gtsima}}           

\newcommand{\aap}{A\&A}
\newcommand{\aj}{AJ}
\newcommand{\apj}{ApJ}
\newcommand{\apjl}{ApJL}
\newcommand{\apjs}{ApJS}
\newcommand{\apss}{ApSS}
\newcommand{\araa}{ARAA}
\newcommand{\mnras}{MNRAS}
\newcommand{\pasj}{PASJ}
\newcommand{\pasp}{PASP}
\newcommand{\physrep}{Phys. Rep.}
\newcommand{\pre}{Phys.\ Rev.\ E}

\newcommand{\BB}{{\bf B}}
\newcommand{\barr}{\begin{eqnarray}}
\newcommand{\beq}{\begin{equation}}
\newcommand{\cs} {c_{\rm s}}
\newcommand{\earr}{\end{eqnarray}}
\newcommand{\eeq}{\end{equation}}
\newcommand{\gamac} {{\gamma_{\rm c}}}
\newcommand{\gamam} {{\gamma_{\rm m}}}
\newcommand{\gamef} {{\gamma_{\rm e}}}
\newcommand{\kms}{~{\rm km~s}^{-1}}

\newcommand{\lambl}{\lambda_{\ell}}
\newcommand{\lambo}{\lambda_0}
\newcommand{\Lc}{L_{\rm c}}

\newcommand{\Ma}{M_{\rm A}}
\newcommand{\mh} {m_{\rm H}}
\newcommand{\Mmol}{M_{\rm mol}}
\newcommand{\Ms}{M_{\rm s}}
\newcommand{\Msun}{M_\odot}
\newcommand{\mum}{\mu_{\rm m}}
\newcommand{\nt}{n_{\rm t}}
\newcommand{\pcc} {{\rm ~cm}^{-3}}
\newcommand{\Peq} {P_{\rm eq}}
\newcommand{\psc} {{\rm ~cm}^{-2}}
\newcommand{\Rey} {R_{\rm e}}
\newcommand{\tcool} {\tau_{\rm c}}
\newcommand{\tff} {\tau_{\rm ff}}
\newcommand{\tturb} {\tau_{\rm t}}
\newcommand{\uu}{{\bf u}}
\newcommand{\va}{\upsilon_{\rm A}}

\makeindex             

\begin{document}

\title*{Interstellar MHD Turbulence and Star Formation}
\author{Enrique V\'azquez-Semadeni}
\institute{Enrique V\'azquez-Semadeni \at Centro de Radioastronom\' \i a
y Astrof\' \i sica, 
Universidad Nacional Aut\'onoma de M\'exico, Campus Morelia, P.O.\ Boz
3-72 (Xangari), Morelia, Michoac\'an, 58089, M\'exico.
\email{e.vazquez@crya.unam.mx}}
%
%
\maketitle

\abstract*{This chapter reviews the nature of turbulence in the Galactic
interstellar medium (ISM) and its connections to the star formation (SF)
process. The ISM is turbulent, magnetized, self-gravitating, and is
subject to heating and cooling processes that control its thermodynamic
behavior, causing it to behave approximately isobarically, in spite of
spanning several orders of magnitude in density and temperature. The
turbulence in the warm and hot ionized components of the ISM appears to
be trans- or subsonic, and thus to behave nearly incompressibly.  However,
the neutral warm and cold components are highly compressible, as a
consequence of both thermal instability (TI) in the atomic gas and of
moderately-to-strongly supersonic motions in the roughly isothermal cold
atomic and molecular components. Within this context, we discuss: i) the
production and statistical distribution of turbulent density
fluctuations in both isothermal and polytropic media; ii) the nature of
the clumps produced by TI, noting that, contrary to
classical ideas, they in general accrete mass from their environment in
spite of exhibiting sharp discontinuities at their boundaries; iii) the
density-magnetic field correlation (and, at low densities, lack thereof)
in turbulent density fluctuations, as a consequence of the superposition
of the different wave modes in the turbulent flow; iv) the evolution of
the mass-to-magnetic flux ratio (MFR) in density fluctuations as they
are built up by dynamic compressions; v) the formation of cold, dense
clouds aided by TI, in both the hydrodynamic and the
magnetohydrodynamic (MHD) cases; vi) the expectation that star-forming
molecular clouds are likely to be undergoing global gravitational
contraction, rather than being near equilibrium, as generally believed, and
vii) the regulation of the star formation rate (SFR) in such
gravitationally contracting clouds by stellar feedback which, rather
than keeping the clouds from collapsing, evaporates and disperses them
while they collapse. }

\abstract{This chapter reviews the nature of turbulence in the Galactic
interstellar medium (ISM) and its connections to the star formation (SF)
process. The ISM is turbulent, magnetized, self-gravitating, and is
subject to heating and cooling processes that control its thermodynamic
behavior, causing it to behave approximately isobarically, in spite of
spanning several orders of magnitude in density and temperature. The
turbulence in the warm and hot ionized components of the ISM appears to
be trans- or subsonic, and thus to behave nearly incompressibly.  However,
the neutral warm and cold components are highly compressible, as a
consequence of both thermal instability (TI) in the atomic gas and of
moderately-to-strongly supersonic motions in the roughly isothermal cold
atomic and molecular components. Within this context, we discuss: i) the
production and statistical distribution of turbulent density
fluctuations in both isothermal and polytropic media; ii) the nature of
the clumps produced by TI, noting that, contrary to
classical ideas, they in general accrete mass from their environment in
spite of exhibiting sharp discontinuities at their boundaries; iii) the
density-magnetic field correlation (and, at low densities, lack thereof)
in turbulent density fluctuations, as a consequence of the superposition
of the different wave modes in the turbulent flow; iv) the evolution of
the mass-to-magnetic flux ratio (MFR) in density fluctuations as they
are built up by dynamic compressions; v) the formation of cold, dense
clouds aided by TI, in both the hydrodynamic (HD) and
the magnetohydrodynamic (MHD) cases; vi) the expectation that
star-forming molecular clouds are likely to be undergoing global
gravitational contraction, rather than being near equilibrium, as generally
believed, and vii) the regulation of the star formation rate (SFR) in
such gravitationally contracting clouds by stellar feedback which,
rather than keeping the clouds from collapsing, evaporates and disperses
them while they collapse. }

\section{Introduction}
\label{sec:intro}

The interstellar medium (ISM) of our galaxy (the Milky Way, or simply,
The Galaxy) is mixture of gas, dust, cosmic rays, and magnetic fields
that occupy the volume in-between stars \citep[e.g.,][]{Ferriere01}. The
gasesous component, with a total mass $ \sim 10^{10} \Msun$, may be in
either ionized, neutral atomic or neutral molecular forms, spanning a
huge range of densities and temperatures, from the so-called hot ionized
medium (HIM), with densities $n \sim 10^{-2} \pcc$ and temperatures $T
\sim 10^6$ K, through the warm ionized and neutral (atomic) media (WIM
and WNM, respectively, both with $n \sim 0.3 \pcc$ and $T \sim 10^4$ K) and
the cold neutral (atomic) medium (CNM, $n \sim 30 \pcc$, $T \sim 100$
K), to the {\it giant molecular clouds} (GMCs, $n \gtrsim 100 \pcc$ and
$T \sim 10$--20 K). These span several tens of parsecs across, and, in
turn, contain plenty of substructure, which is commonly classified into
{\it clouds} ($n \sim 10^3 \pcc$, size scales $L$ of a few parsecs),
{\it clumps} ($n \sim 10^4 \pcc$, $L \sim 1$ pc), and {\it cores} ($n
\gtrsim 10^5 \pcc$, $L \sim 0.1$ pc). It is worth noting that the
temperature of most molecular gas is remarkably uniform, $\sim 10$--30
K.

Moreover, the ISM is most certainly turbulent, as typical estimates of
the Reynolds number ($\Rey$) within it are very large. For example, in
the cold ISM, $R_{\rm e} \sim 10^5$--$10^7$ \citep[][\S
4.1]{ES04}. This is mostly due to the very large spatial scales involved
in interstellar flows.  Because the temperature of the ISM varies so much
from one type of region to another, so does the sound speed, and
therefore the turbulent velocity fluctuations are often moderately or
even strongly supersonic \citep[e.g., ][and references therein]{HT03,
ES04}. In these cases, the flow is significantly compressible, inducing
large-amplitude (nonlinear) density fluctuations. The density
enhancements thus formed constitute dense clouds and their substructure
\citep[e.g.,][]{Sasao73, Elm93, BP+99a}.

In addition to being turbulent, the ISM is subject to a number of
additional physical processes, such as gravitational forces exerted by
the stellar and dark matter components as well as by its own
self-gravity, magnetic fields, cooling by radiative microscopic
processes, and radiative heating due both to nearby stellar sources as
well as to diffuse background radiative fields. It is within this
complex and dynamical medium that stars are formed by the gravitational
collapse of certain gas parcels.

In this chapter, we focus on the interaction between turbulence, the
effects of radiative heating and cooling, which effectively enhance the
compressibility of the flow, the self-gravity of the gas, and magnetic
fields. Their complex interactions have a direct effect on the star
formation process. The plan of the chapter is as follows: in \S
\ref{sec:thermodynamics} we briefly recall the effects that the net
heating and cooling have on the effective equation of state of the flow
and, in the case of thermally unstable flows, on its tendency to
spontaneously segregate in distinct phases. Next, in \S
\ref{sec:turb} we discuss a few basic notions about turbulence and the
turbulent production of density fluctuations in both the hydrodynamic
(HD) and magnetohydrodynamic cases, to then discuss, in \S
\ref{sec:turb_therm}, the evolution and properties of clouds and clumps
formed by turbulence in multiphase media. In \S \ref{sec:turb_multi}, we
discuss the likely nature of turbulence in the diffuse (warm and hot)
components of the ISM, as well as in the dense, cold atomic and molecular
clouds, suggesting that in the latter, at least during the process of
forming stars, the velocity field may be dominated by gravitational
contraction. Next, in \S \ref{sec:turb_SF} we discuss the regulation of
star-formation (SF) in gravitationally contracting molecular clouds
(MCs), in particular whether it is accomplished by magnetic support,
turbulence, or stellar feedback, and how. Finally, in \S
\ref{sec:conclusions} we conclude with a summary and some final remarks.

\section{ISM Thermodynamics: Thermal Instability} \label{sec:thermodynamics}

The ISM extends essentially over the entire disk of the Galaxy and, when
considering a certain dense subregion of it, such as a cloud or cloud
complex, it is necessary to realize that any such subregion constitutes
an open system, whose interactions with its environment need to be taken
into account. A fundamental form of interaction with the surroundings,
besides dynamical interactions, is through the exchange of heat, which
occurs mostly through heating by the UV background radiation produced by
distant massive stars, local heating when nearby stellar sources (OB
star ionization heating and supernova explosions), cosmic ray heating,
and cooling by thermal and line emission from dust and gas, respectively
\citep[see, e.g.,][]{DM72, Wolfire+95}.

\begin{figure}
\begin{center}
\includegraphics[scale=.4]{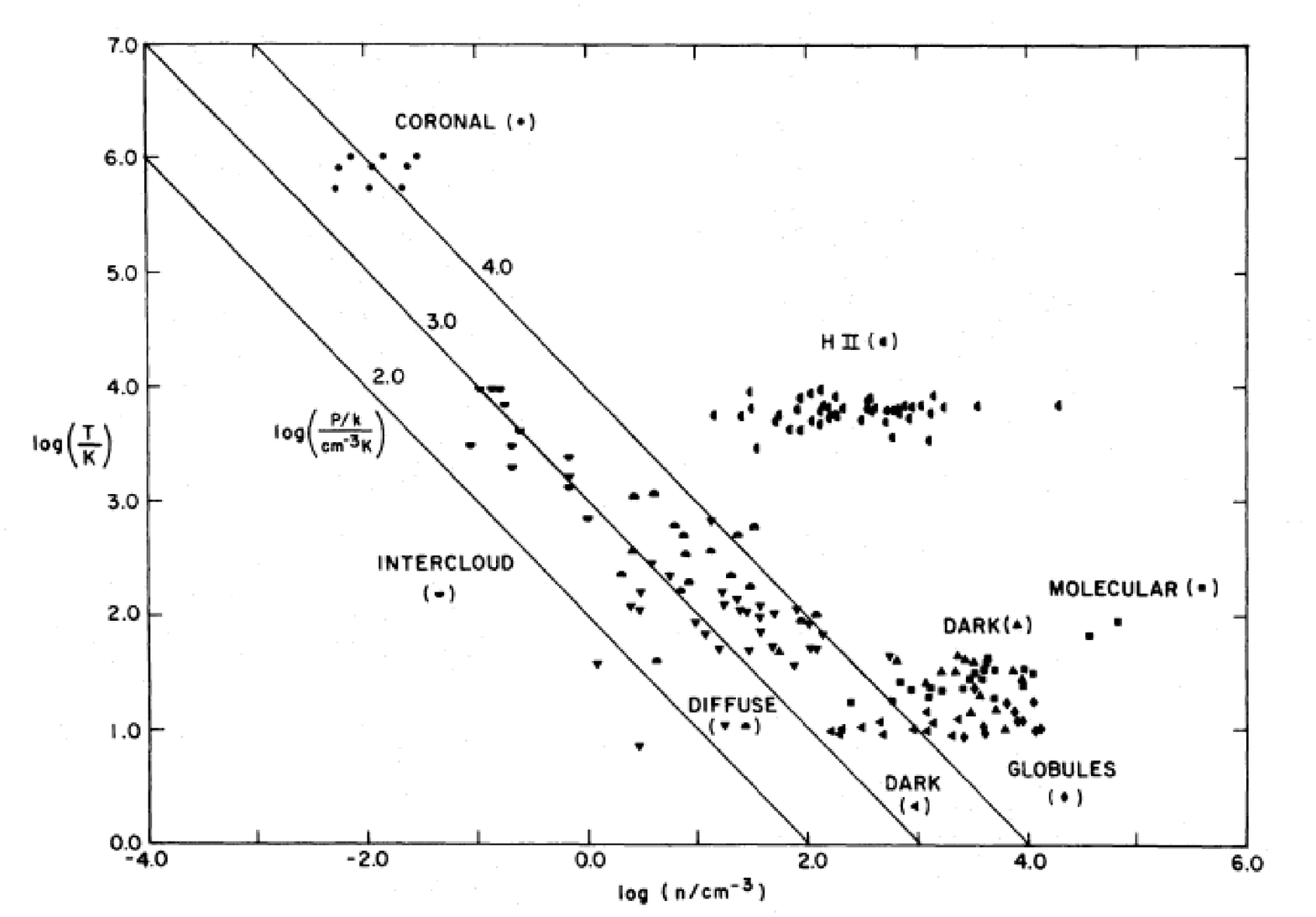}
\end{center}
\caption{Thermal pressure in various types of interstellar
regions. The points labeled {\it coronal} correspond essentially to what
we refer to as the HIM in the text; {\it intercloud} regions refer to
the WIM and WNM; {\it diffuse}, to CNM clouds, and {\it dark, globule}
and {\it molecular} to molecular gas. From \citet{Myers78}. }
\label{fig:Myers78}
\end{figure}

Globally, and as a first approximation, the ISM is roughly isobaric, as
illustrated in Fig.\ \ref{fig:Myers78}. As can be
seen there, most types of regions, either dilute or dense, lie within an
order of magnitude from a thermal pressure $P \sim 3000 $
K$\pcc$. The largest deviations from this
pressure uniformity are found in {\sc Hii} regions, which are the ionized
regions around massive stars due to their UV radiation output, and in
molecular clouds, which, as we shall see in \S \ref{sec:molec_turb},
are probably pressurized by gravitational compression.

\begin{figure}
\begin{center}
\includegraphics[scale=.3]{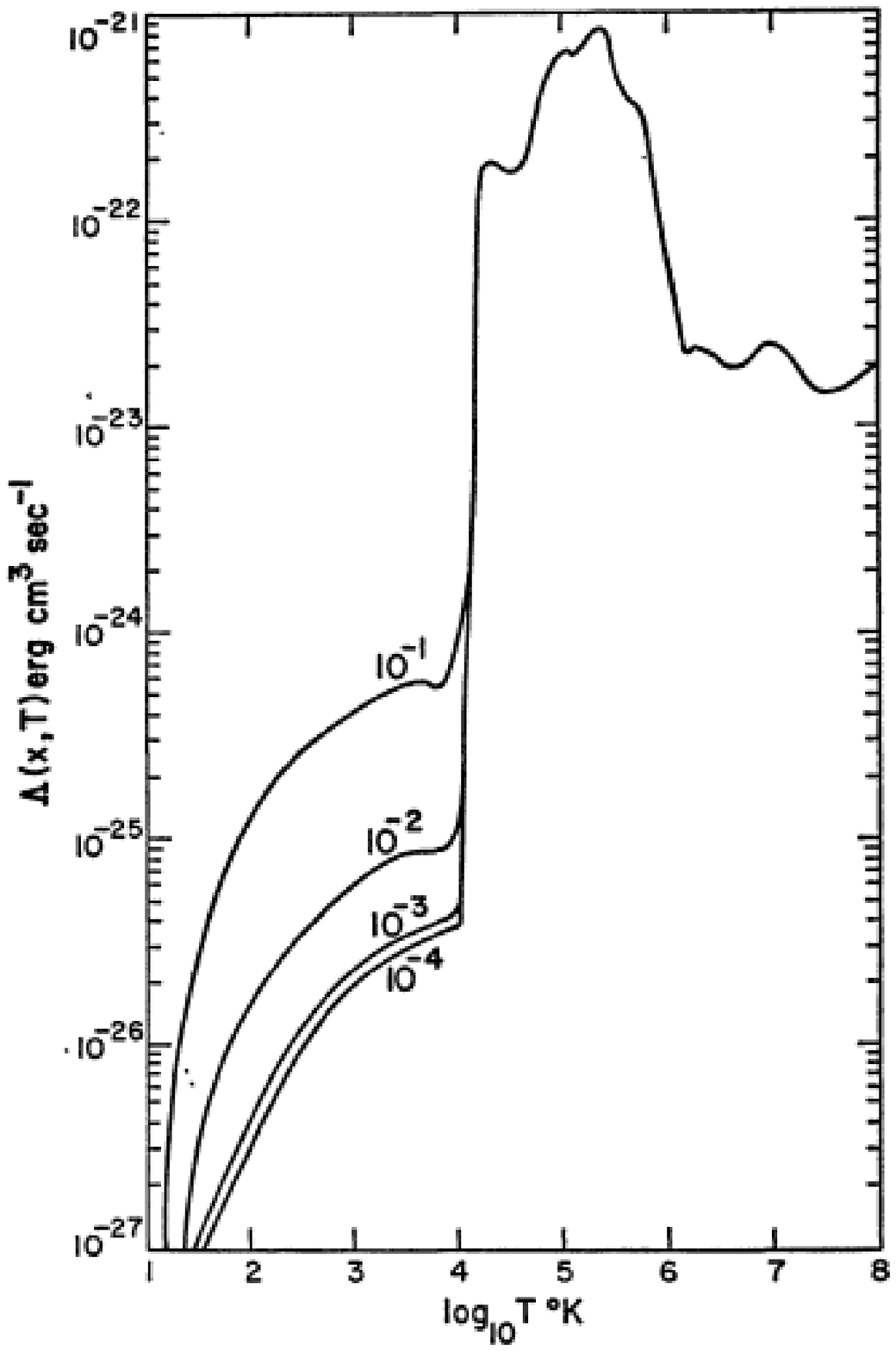}
\includegraphics[scale=.45]{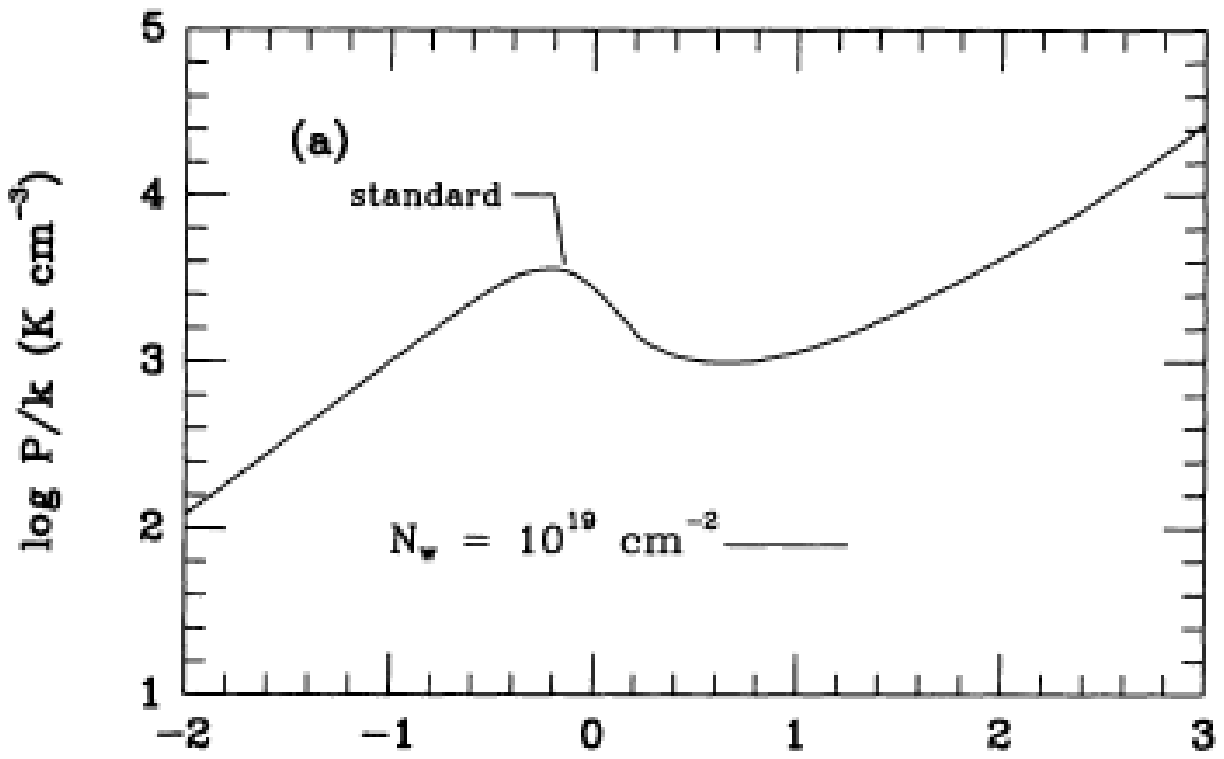}
\end{center}
\caption{{\it Left:} Temperature dependence of the cooling function
$\Lambda$. The labels indicate values of the ionization fraction (per
number) of the gas. From \citet{DM72}. {\it Right:} Thermal-equilibrium
pressure $\Peq$ as a function of number density for ``standard''
conditions of metallicity and background UV radiation for the atomic
medium. The horizontal axis gives $\log_{10}(n/{\rm cm}^3)$. From
\citet{Wolfire+95}. }
\label{fig:DM72_Peq_vs_n}
\end{figure}

The peculiar thermodynamic behavior of the ISM is due to the functional
forms of the radiative heating and cooling functions acting on it, which
depend on the density, temperature, and chemical composition of the gas.
The {\it left} panel of Fig.\ \ref{fig:DM72_Peq_vs_n} shows the temperature
dependence of the cooling function $\Lambda$ \citep{DM72}. One
well-known crucial consequence of this general form of the cooling is
that the atomic medium is {\it thermally unstable} \citep{Field65} in
the density range $1 \lesssim n \lesssim 10 \pcc$ (corresponding to
$5000 \gtrsim T \gtrsim 30$ K), meaning that the medium tends to
spontaneously segregate into two stable {\it phases}, one warm and diffuse,
with $n \sim 0.3 \pcc$ and $T \sim 8000$ K, and the other cold and
dense, with $n \sim 30 \pcc$ and $T \sim 80$ K, both at a pressure $P/k
\sim 2500 {~\rm K} \pcc$ \citep[][see also the reviews by Meerson 1996; 
V\'azquez-Semadeni et al. 2003; V\'azquez-Semadeni 2013]{FGH69,
Wolfire+95}, as illustrated in the {\it right} panel of Fig.\
\ref{fig:DM72_Peq_vs_n}. The cold gas is expected to form small clumps,
since the fastest growing mode of the instability occurs at vanishingly
small scales in the absence of thermal conductivity, or at scales $\sim
0.1$ pc for the estimated thermal conductivity of the ISM
\citep[see, e.g.,][]{Field65, AH05}. Because the atomic gas in the ISM
has two stable phases, it is often referred to as a {\it thermally
bistable} medium.

It is important to note that, even if the medium is {\it not} thermally
unstable, the balance between heating and cooling implies a certain
functional dependence of $\Peq(\rho)$, which is often approximated by a
{\it polytropic} law of the form $\Peq \propto \rho^\gamef$
\citep[e.g.,][]{Elm91, VPP96}, where $\gamef$ is the {\it effective
polytropic exponent}. In general, $\gamef$ is {\it not} the ratio of
specific heats for the gas in this case, but rather a free parameter
that depends on the functional forms of $\Lambda$ and $\Gamma$. The
isobaric mode of thermal instability (TI) corresponds to $\gamef < 0$. A
flow is sometimes said to be {\it softer} as the parameter $\gamef$
becomes smaller.

\section{Compressible Polytropic MHD Turbulence} \label{sec:turb}

\subsection{Equations} \label{sec:eqs}

In the previous section we have discussed thermal aspects of the ISM,
whose main dynamical effect is the segregation of the medium into the
cold and warm phases. Let us now discuss dynamics. As was mentioned in
\S \ref{sec:intro}, the ISM is in general turbulent and magnetized,
and therefore it is necessary to understand the interplay between
turbulence, magnetic fields, and the effects of the net cooling ($n
\Lambda - \Gamma$), which affects the compressibility of the gas
\citep{VPP96}.

The dynamics of the ISM are governed by the fluid equations,
complemented by self-gravity, the heating and cooling terms in the
energy equation, and the equation of magnetic flux conservation
\citep[e.g.,][]{Shu92}:
\barr
\frac{\partial\rho}{\partial t} + \uu \cdot \nabla \rho &=& -\rho \nabla
\cdot \uu, \label{eq:contin}\\
\frac{\partial\uu}{\partial t} +  \uu\cdot\nabla\uu &=&
-\frac{\nabla P}{\rho} -  \nabla \varphi + \nu \left[\nabla^2 \uu +
\frac{\nabla (\nabla \cdot \uu)}{3} \right]
+ \frac{1}{4 \pi \rho} \bigl(\nabla
\times {\BB}\bigr) \times {\BB},  \label{eq:mom_cons} \\
\frac{\partial e}{\partial t} + \uu \cdot \nabla e &=& -(\gamma - 1) e
\nabla \cdot \uu  + \Gamma - n \Lambda,
\label{eq:int_en_cons} \\
\frac{\partial {\vec B}}{\partial t} + \nabla \times (\vec B \times \uu)
&=& - \nabla \times (\eta \nabla \times \BB) + \nabla \times
\left\{\frac{\BB}{4 \pi \gamac \rho_{\rm n} \rho_{\rm i}} \times
\left[\BB \times \left(\nabla \times \BB \right) \right] \right\},
\label{eq:flux_freez} \\ 
\nabla^2 \varphi &=& 4 \pi G \rho, \label{eq:Poisson}
\earr
where $\rho=\mum \mh n$ is the mass density, $\mum$ is the mean particle
mass, $\mh$ is the hydrogen mass, $\uu$ is the velocity vector, $e$ is
the internal energy per unit mass, $\BB$ is the magnetic field strength
vector, $\varphi$ is the gravitational potential, $\nu$ is the kinematic
viscosity, $\eta$ is the electrical resistivity, and $\gamac$ is the
collisional coupling constant between neutrals and ions in a partially
ionized medium. Equation (\ref{eq:contin}) represents mass conservation,
and is also known as the {\it continuity equation}. Equation
(\ref{eq:mom_cons}) is the momentum conservation, or {\it Navier-Stokes}
equation per unit mass, with additional source terms representing the
gravitational force $\nabla
\varphi/\rho$ and the Lorentz force. In turn, the gravitational
potential is given by {\it Poisson's equation}, eq.\
(\ref{eq:Poisson}). Equation (\ref{eq:int_en_cons}) represents the
conservation of internal energy, with $\Gamma$ being the heating
function and $\Lambda$ the cooling function. The combination $n\Lambda -
\Gamma$ is usually referred to as the {\it net cooling}, and the
condition $n\Lambda = \Gamma$ is known as the {\it thermal equilibrium}
condition. Finally, eq.\ (\ref{eq:flux_freez}) represents the
conservation of magnetic flux (see below). Equations
(\ref{eq:contin})--(\ref{eq:Poisson}) are to be solved simultaneously,
given some initial and boundary conditions.

A brief discussion of the various terms in the momentum and
flux conservation equations is in order. In eq.\ (\ref{eq:mom_cons}), the
second term on the left is known as the {\it advective} term, and
represents the transport of $i$-momentum by the $j$ component of the
velocity, where $i$ and $j$ represent any two components of the
velocity. It is responsible for {\it mixing}. The pressure gradient term
(first term on the right-hand side [RHS]) in general acts to counteract
pressure, and therefore density, gradients across the flow. The
term in the brackets on the RHS, the {\it viscous} term, being of a
diffusive nature, tends to erase velocity gradients, thus tending to
produce a uniform flow. Finally, the last term on the RHS is
the Lorentz force.

On the other hand, eq.\ (\ref{eq:flux_freez}), assuming $\eta = 0$
(i.e., zero electrical resistivity, or equivalently, infinite
conductivity) and $\gamac \rightarrow \infty$ (i.e., perfect coupling
between neutrals and ions), implies that the magnetic flux $\Phi$ through a
Lagrangian cross-sectional area $A$, given by
\beq
\Phi \equiv \int_A \BB \cdot \hat{\vec n}~\D A,
\label{eq:B_flux}
\eeq
remains constant in time as the area moves with the flow. This is the
property known as {\it flux 
freezing}, and implies that the gas can slide freely along field lines,
but drags the field lines with it when it moves perpendicularly to
them. Note that this condition is often over-interpreted as to imply that
the magnetic and the density fields must be correlated, but this is an
erroneous notion. Only motions perpendicular to the magnetic field lines
produce a correlation between the two fields, while motions parallel to
the lines leave the magnetic field unaffected, while the density field
can fluctuate freely. We discuss this at more length in \S
\ref{sec:dens_fluc_mag}. 

The first term on the RHS of eq.\ (\ref{eq:flux_freez})
represents dissipation of the magnetic flux by electrical resistivity,
and gives rise to the phenomenon of {\it reconnection} of field lines
(see, e.g., the book by Shu 1992, and the review by Lazarian 2012).  The
second term on the RHS of eq.\ (\ref{eq:flux_freez}) represents {\it
ambipolar diffusion} (AD), the deviation from the perfect flux-freezing
condition that occurs for the neutral particles in the flow due to their
slippage with respect to the ions in a partially ionized medium. We will
further discuss the role of AD in the process of star formation (SF)
in \S\S \ref{sec:MFR_evol} and \ref{sec:molec_turb_magn}.

\subsection{Governing Non-Dimensional Parameters} \label{sec:Re_Ms}

Turbulence develops in a flow when the ratio of the advective term to
the viscous term becomes very large. That is,
\beq
\frac{{\cal O}\left[\uu\cdot\nabla\uu \right]} {{\cal O} \left[\nu
\left(\nabla^2 \uu + \frac{\nabla (\nabla \cdot \uu)}{3} \right) \right]} \sim
\frac{U^2}{L} \left[\nu \frac{U}{L^2}\right]^{-1} \sim \frac{UL}{\nu}
\equiv \Rey \gg 1,
\label{eq:Reynolds_num}
\eeq
where $\Rey$ is the {\it Reynolds number}, $U$ and $L$ are
characteristic velocity and length scales for the flow, and ${\cal O}$
denotes ``order of magnitude''. This condition
implies that the mixing action of the advective term overwhelms the
velocity-smoothing action of the viscous term.

On the other hand, noting that the advective and pressure gradient
terms contribute comparably to the production of density fluctuations, we
can write
\barr
1 \sim \frac{{\cal O}\left(\uu\cdot\nabla\uu \right)} {{\cal O}\left(\nabla P/\rho \right)} \sim \frac{U^2}{L}
\left[\frac{\Delta P}{L \rho} \right]^{-1} &\sim& U^2 \left(\frac{\cs^2
\Delta \rho}{\rho} \right)^{-1} \equiv \Ms^2 \left(\frac{\Delta \rho}
{\rho} \right)^{-1},\\
\Rightarrow \frac{\Delta \rho}{\rho} &\sim& \Ms^2 \label{eq:dens_jumps},
\earr
where $\Ms \equiv U/\cs$ is the {\it sonic Mach number}, $\Delta
\rho/\rho$ is the {\it density jump}, and we have
made the approximation that $\Delta P/\Delta \rho \sim \cs^2$, where
$\cs$ is the sound speed. Equation (\ref{eq:dens_jumps}) then implies
that strong compressibility requires $\Ms \gg 1$. Conversely, flows with
$\Ms \ll 1$ behave incompressibly, even if they are gaseous. Such is the
case, for example, of the Earth's atmosphere. In the incompressible
limit, $\rho =$ cst., and thus eq.\ (\ref{eq:contin}) reduces to $\nabla
\cdot \uu = 0$. Note, however, that the requirement $\Ms \gg 1$ for
strong compressibility applies for flows that behave nearly
isothermally, for which the approximation $\Delta P/\Delta \rho \sim
\cs^2$ is valid, while ``softer'' (cf.\ \S \ref{sec:thermodynamics}) flows
have much larger density jumps at a given Mach number. For example,
\citet{VPP96} showed that 
polytropic flows of the form $P \propto \rho^\gamef$ with $\gamef
\rightarrow 0$, have density jumps of the order $e^{\Ms^2}$.

A trivial, but often overlooked, fact is that, in order to
produce a density enhancement in a certain region of the flow, the
velocity at that point must have a negative divergence (i.e., a {\it
convergence}), as can be seen by rewriting eq.\
(\ref{eq:contin}) as
\beq
\frac{\D \rho}{\D t} = - \rho \nabla \cdot \uu,
\label{eq:cont2}
\eeq
where $\D/\D t \equiv \partial/\partial t +\uu \cdot \nabla$ is the {\it
total}, {\it material}, or {\it Lagrangian} derivative. However, it is
quite common to encounter in the literature discussions of pre-existing
density enhancements (``clumps'') in hydrostatic equilibrium. It
should be kept in mind that these can only exist in multi-phase media,
where a dilute, warm phase can have the same pressure as a denser, but
colder, clump. But even in this case, the {\it formation} of that clump
must have initially involved the convergence of the flow towards the
cloud, and the hydrostatic situation is applicable in the limit of very
long times after the formation of the clump, when the convergence of the
flow has subsided.

Finally, two other important parameters determining the properties of a
magnetized flow are the {\it Alfv\'enic Mach number}, $\Ma \equiv U/\va$
and the {\it plasma beta}, $\beta \equiv P_{\rm th}/P_{\rm mag}$
where $\va = B/\sqrt{4 \pi \rho}$ is the {\it Alfv\'en
speed}. 
Similarly to the non-magnetic case, large values of the
Alfv\'enic Mach number are required in order to produce significant
density fluctuations through compressions {\it perpendicular} to the
magnetic field. However, it is important to note, as mentioned in \S
\ref{sec:eqs}, that compressions {\it along} the magnetic field
lines are not opposed at all by magnetic forces. Note that, in the
isothermal case, $\beta = 2 \cs^2/\va^2$.

\subsection{Production of Density Fluctuations. The Non-Magnetic Case}
\label{sec:dens_fluc_nonmag}  

As mentioned in the previous sections, strongly supersonic motions, or
the ability to cool rapidly, allow the production of large-amplitude
density fluctuations in the flow. Note, however, that the nature of
turbulent density fluctuations in a single-phase
medium\footnote{Thermodynamically, a {\it phase} is a region of space
throughout which all physical properties of a material are essentially
uniform \citep[e.g.,][]{MR74}. A {\it phase transition} is a boundary
that separates physically distinct phases, which differ in most
thermodynamic variables except one (often the pressure). See
\citet{VS09} for a discussion on the nature of phases and phase
transitions in the ISM.} (such as, for
example, a regular isothermal or adiabatic flow) is very different from
that of the cloudlets formed by TI (cf.\ \S
\ref{sec:thermodynamics}). In a single-phase
turbulent medium, turbulent density fluctuations must be transient,
because in this case a higher density generally implies a higher
pressure,\footnote{An exception would be a so-called Burgers' flow,
which is characterized by the absence of the pressure gradient term
\citep{Burgers74}, and can be thought of as the transitional regime from
thermal stability to instability.} and therefore the fluctuations must
re-expand (in roughly a sound-crossing time) after the compression that
produced them has subsided.

Note that the above result includes the case with self-gravity, since in
single-phase media, although hydrostatic equilibrium solutions do exist,
they are generally unstable. Specifically, the singular isothermal
sphere is known to be unstable \citep{Shu77}, and non-singular
configurations such as the Bonnor-Ebert (BE) spheres \citep{Ebert55,
Bonnor56} need to be truncated so that the central-to-peripheral density
ratio is smaller than a critical value $\sim 14$ in order to be
stable. Such stable configurations, however, need to be confined by some
means to prevent their expansion. Generally, the confining agent is
assumed to consist of a dilute, warm phase that provides pressure
without adding additional weight. However, such a warm phase is not
available in single-phase flows, and the only way to confine the BE
sphere is to continuously extend it to infinity, in which case the
central-to-peripheral density ratio also tends to infinity, and the
configuration is unstable \citep{VS+05a}.

Instead, in {\it multiphase} flows, abrupt density variations may exist
between different phases even though they may be at roughly the same
thermal pressure, and therefore, the dense clumps do not tend to
re-expand. In the remainder of this section we discuss the probability
distribution of the density fluctuations, the nature of the resulting
clumps and their interfaces with their environment, the correlation
between the magnetic and density fields, and the evolution of the
mass-to-magnetic flux ratio as the clumps are assembled by turbulent
compressions.

\subsubsection{The Probability Distribution of Density Fluctuations. The
Non-Magnetic Case } \label{sec:dens_PDF} 

For astrophysical purposes it is important to determine the distribution
of the density fluctuations, as they may constitute, or at least provide
the seeds for, what we normally refer to as ``clouds'' in the ISM. In
single-phase media, however, due to the transient nature of turbulent
density fluctuations, this distribution refers to a time-stationary
population of fluctuations, even though the fluctuations themselves will
appear and disappear on timescales of the order of their crossing time at
the speed of the velocity fluctuations that produce them.

The probability density function (PDF) of the density field in turbulent
isothermal flows was initially investigated through numerical
simulations.  \citet{VS94} found that, in the isothermal case, the PDF
posesses a lognormal form.  A theory for the emergence of this
functional form was later proposed by \citet{PV98}, in which the
production of density fluctuations was assumed to arise from a
succession of random compressive or expansive waves, each one acting on
the value of the density left by the previous one.  The amplitude of
each wave can then be described as a random variable, characterized by
some probability distribution. Because the medium contains a unique
distribution of (compressible) velocity fluctuations, and because the
density jumps in isothermal flow depend only on Mach number but not on
the local density, the density fluctuations belong all to a unique
distribution as well, yet each one can be considered independent of the
others if the global time scales considered are much longer than the
autocorrelation time of the velocity divergence
\citep{Blaisdell+93}. Finally, because the density jumps are
multiplicative in the density (cf.\ eq.\ \ref{eq:dens_jumps}), then they
are additive in $s \equiv \ln \rho$. Under these conditions, the Central
Limit Theorem can be invoked for the increments in $s$, implying that
$s$ will be normally distributed, independently of the distribution of
the waves. In consequence, $\rho$ will have a lognormal PDF.

\begin{figure}
\begin{center}
\includegraphics[scale=.32]{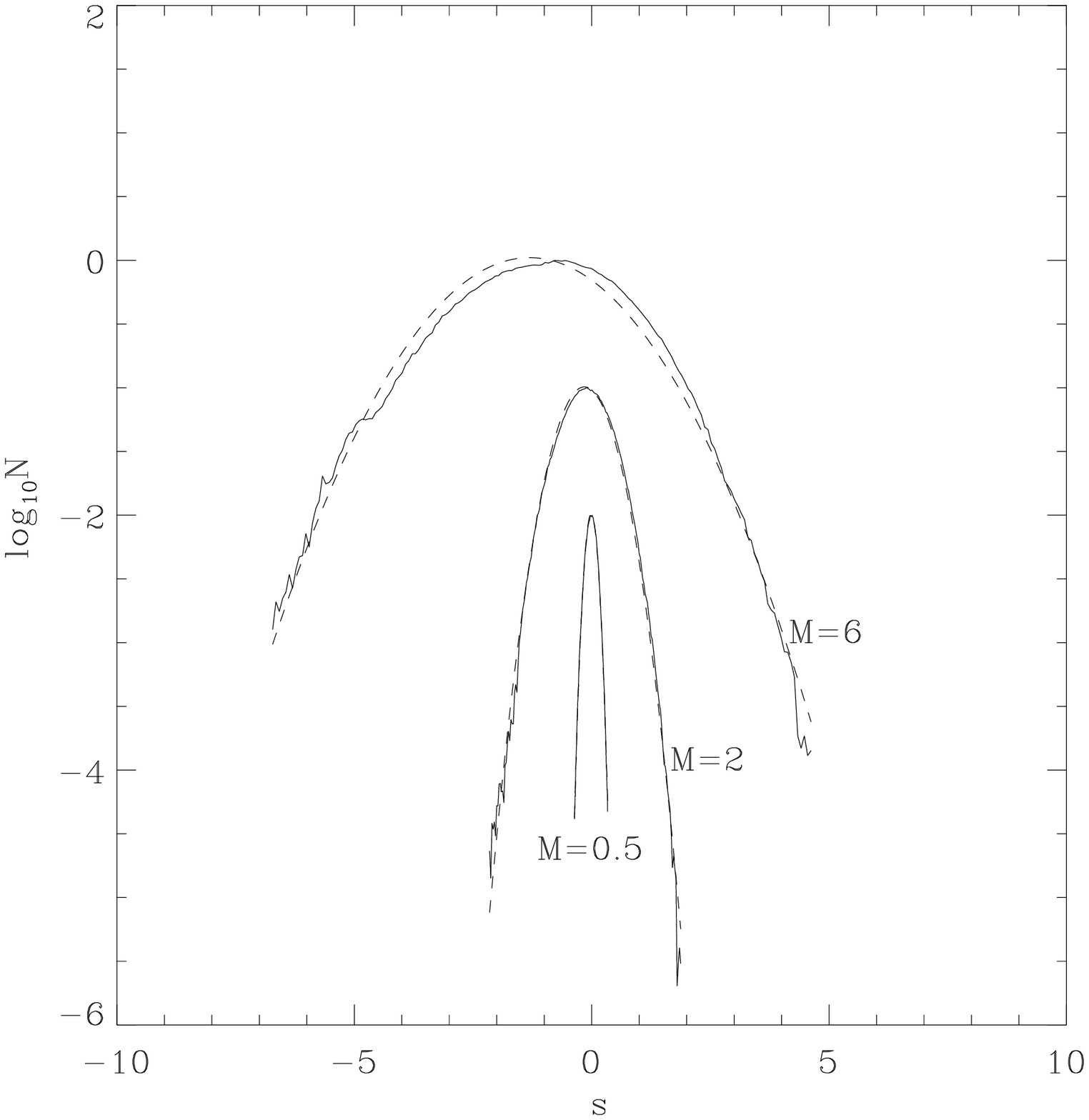}
\includegraphics[scale=.25]{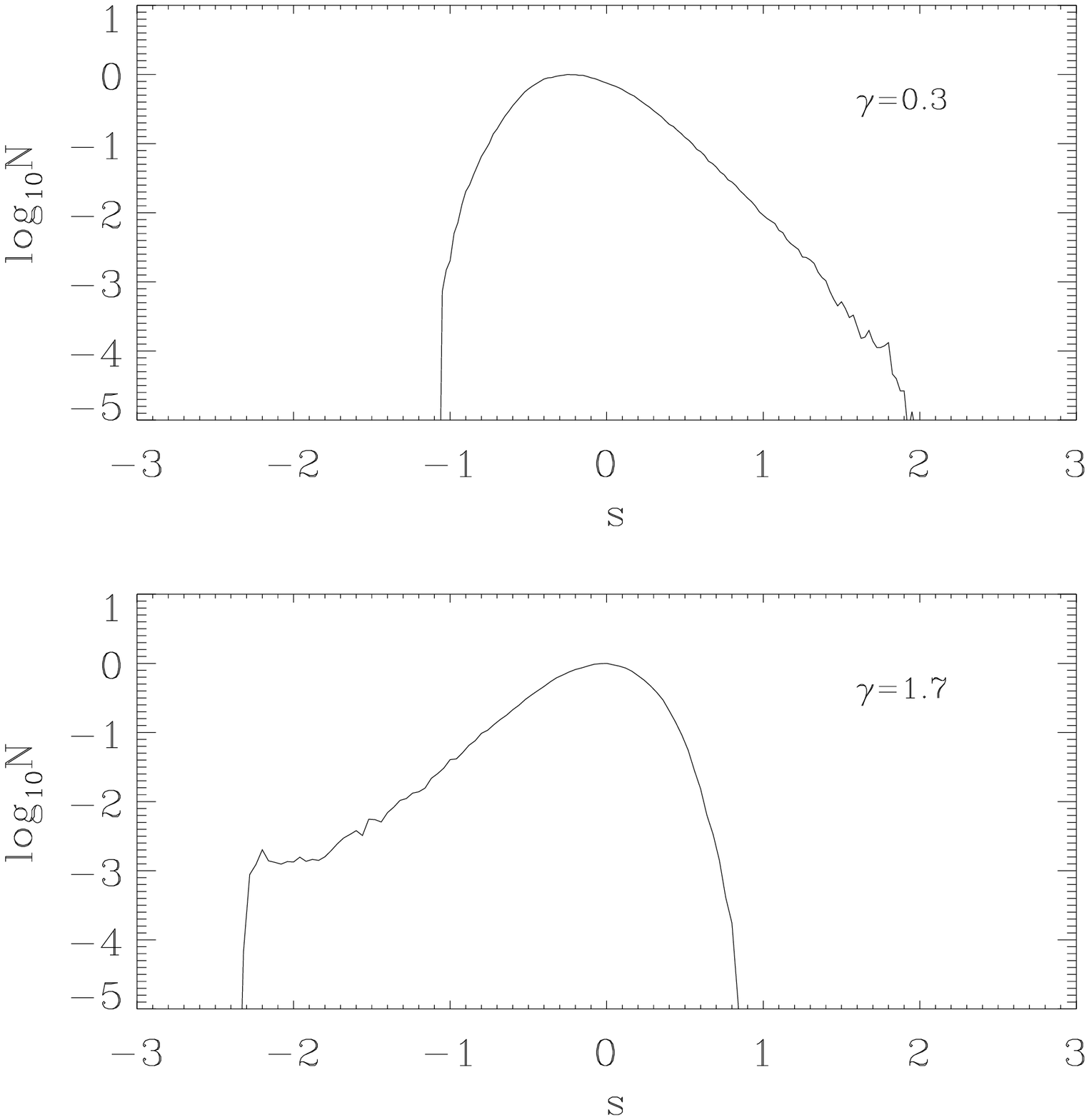}
\end{center}
\caption{{\it Left:} Lognormal density PDFs for isothermal
one-dimensional simulations at various Mach numbers, indicated by the
labels. The independent variable is $s \equiv \ln \rho$. {\it Right:}
Density PDFs for polytropic cases (i.e., with $P \propto \rho^{\gamef}$),
with effective polytropic exponent 
$\gamef = 0.3$ ({\it top}) and $\gamef = 1.7$ ({\it bottom}). From
\citet{PV98}.}
\label{fig:pdfs_PV98}
\end{figure}

In addition, \citet{PV98} also argued that the variance of the density
fluctuations should scale linearly with $\Ms$, a suggestion that has been
investigated further by other groups \citep{Padoan+97,
Federrath+08}. In particular, using numerical simulations of
compressible turbulence driven by either solenoidal (or ``vortical'') or
compressible (or ``potential'') forces, the latter authors proposed that the
variance of $s$ is given by 
\beq 
\sigma_s = \ln(1 + b \Ms^2),
\label{eq:sigma_Ms}
\eeq
where $b$ is a constant whose value depends on the nature of the forcing,
taking the extreme values of $b=1/3$ for purely solenoidal forcing, and
$b=1$ for purely compressible forcing. The lognormal density PDF for
the one-dimensional, isothermal simulations of \citet{PV98}, with
its dependence on $\Ms$, is illustrated in the {\it left} panel of Fig.\
\ref{fig:pdfs_PV98}. 
\begin{figure}
\begin{center}
\includegraphics[scale=.5]{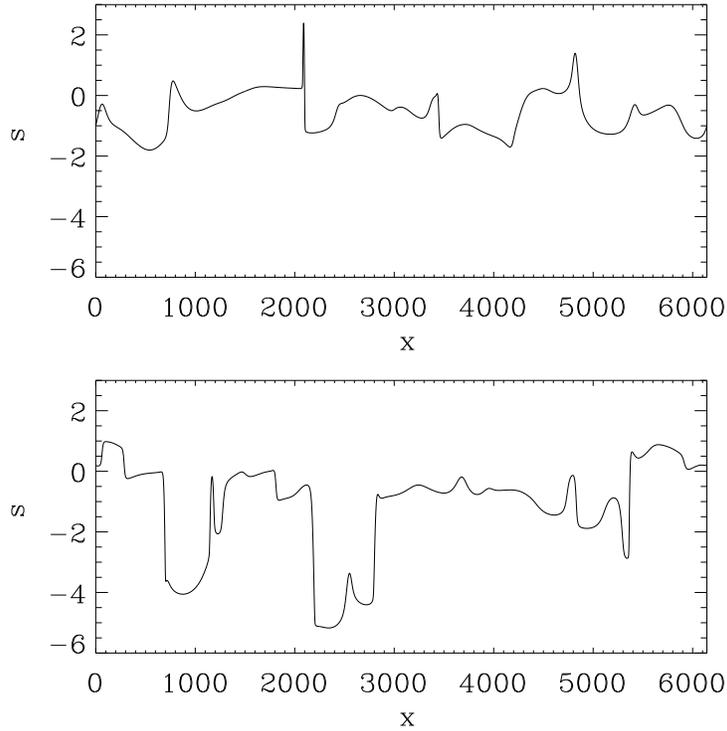}
\end{center}
\caption{Plots of $s = \ln \rho$ of the density field in
one-dimensional simulations of polytropic flows with a resolution of
6144 grid points by \citet{PV98}. {\it Top:} A simulation with $\gamef =
0.5$, exhibiting high-amplitude, narrow density peaks, due to the low
sound speed at high density. {\it Bottom:} A simulation with $\gamef =
1.5$, exhibiting low-amplitude, extended density peaks and deep ``voids''
due to the high values of the sound speed at high densities, and low
values at low densities.}
\label{fig:rho_PV98}
\end{figure}
Finally, \citet[][see also Padoan \& Nordlund 1999]{PV98} also
investigated the case where the flow behaves as a polytrope (cf.\ \S
\ref{sec:thermodynamics}) with
arbitrary values of $\gamef$, by noting that in this case the sound
speed is not constant, but rather depends on the density as $\cs \propto
\rho^{(\gamef -1)/2}$, implying that the local Mach number of a fluid
parcel now depends on the local density besides its dependence on the
value of the flow velocity. Introducing this dependence of $\Ms$ on
$\rho$ in the expression for the lognormal PDF, \citet{PV98} concluded
that the density PDF should develop a power-law tail, at high densities
when $\gamef <1$, and at low densities when $\gamef > 1$. This result
was then confirmed by numerical simulations of polytropic turbulent
flows (Fig.\ \ref{fig:pdfs_PV98}, {\it right} panel). Physically, the
cause for the deviation of the PDF from the lognormal shape is that, for
$\gamef > 1$, the sound speed increases with increasing density, and
therefore, high-density regions can re-expand and disappear quickly,
while ``voids'', with a lower sound speed, last for long times. For
$\gamef < 1$, the sound speed {\it decreases} for increasing density,
and the behavior is reversed: large-amplitude density peaks have lower
sound speeds and therefore last for longer times, while the voids have
higher sound speeds and disappear quickly. The resulting topology of the
density field is illustrated in the one-dimensional case in Fig.\
\ref{fig:rho_PV98}.

\subsubsection{The Nature of Turbulent Clumps} \label{sec:clump_nature}

\paragraph{The Ambiguity of Clump Boundaries and Masses}

The clumps produced as turbulent density fluctuations are precisely
that: fluctuations in a continuum. Besides, there is in general a mass
and energy flux through any fixed boundary we choose to define around
the local density maximum \citep{BP+99a}. This is especially true in
isothermal flows, where no transition from a diffuse phase to a dense
one with the same pressure can occur, so that the only density
discontinuities possible are those produced by shocks. In this case, the
density fluctuations produced by turbulent compressions (a transient
event of elevated ram pressure\footnote{By ``ram'' or ``hydrodynamic''
pressure, we refer to the pressure exerted by the coherent motion, at
speed $\upsilon$, of a fluid of density $\rho$, and is given by $\rho
\upsilon^2$. A familiar example of this is the pressure exerted by a
water jet coming out of a fireman's hose}) must always be transients, and must
eventually re-expand or collapse (see \S \ref{sec:dens_fluc_nonmag}). In
thermally bistable media, the ``boundaries'' are somewhat better
defined, although there is still mass flux through them (see \S
\ref{sec:cloud_form_hd}).

The elusiveness of the notion of clump boundaries implies that we must
rethink some of our dearest notions about clumps. First of all, the mass
of a cloud or clump is not well defined, and additionally must evolve in
time. It is ill-defined because the clump boundary itself is. Several
procedures exist for ``extracting'' clumps from observational maps or
from numerical simulations. One of the most widely employed algorithms
for locating clumps is {\it Clumpfind} \citep{Williams+94}, which works
by locating local peaks in the field being examined and then following
the field down its gradient until another clump profile is met, at which
point an arbitrary boundary is defined between the two. However, this
procedure, by construction, is uncapable of recognizing
``hierarchical'', or ``nested'', structures, where one coherent
``parent'' clump contains other equally coherent ``child''
ones. Moreover, not surprisingly, it has been shown that the clump sets
obtained from application of this algorithm depend sensitively on the
parameters chosen for the definition of the clumps
\citep{Pineda+09}. Conversely, a technique that, by definition, {\it is}
capable of detecting ``parent'' structures, is based on ``structure
trees'', or ``dendrograms'' \citep[e.g.,][]{HS92, Rosolowsky+08}, which
works by thresholding an image at successive intensity levels, and
following the ``parent''-``child'' relationship between the
structures identified at the different levels. Clearly, the two
techiques applied to the same data produce very different sets of clumps
and, in consequence, different clump mass distributions.

This variety of procedures for defining clumps illustrates the ambiguity
inherent in defining a finite {\it object} that is actually part of a
continuum, and implies that the very concept of the mass of a clump
carries with it a certain level of inherent uncertainty.

\paragraph{Clump masses evolve in time}

A turbulent density fluctuation is a local density enhancement produced
by a velocity field that at some moment in time is locally convergent,
as indicated by eq.\ (\ref{eq:cont2}). This process accumulates mass in
a certain region of space (``the clump''), with the natural consequence
that the mass of the clump must increase with time, at least initially, if
the clump is defined, for example, as a connected object with density
above a certain threshold. This
definition corresponds, for example, to clumps defined as compact
objects observed in a particular molecular tracer, since such tracers
require the density to be above a certain threshold to be excited.

The growth of the clump's density (and mass) lasts as long as the total
pressure within the clump (which may include thermal, turbulent and
magnetic components) is smaller than the ram pressure from the
compression. However, in a single-phase medium, once the turbulent
compression subsides, the clump, which is at higher density than its
surroundings, and therefore also at a higher pressure, must therefore
begin to re-expand, unless it manages to become gravitationally unstable
and proceed to collapse \citep[][see also \S
\ref{sec:dens_fluc_nonmag}]{VS+05a, Gomez+07}. Therefore, the mass 
above the clump-defining 
density threshold may begin to decrease again. (Again, see \S
\ref{sec:cloud_form_hd} for the case of clumps forming in multi-phase
media.) As we shall see in \S \ref{sec:MFR_evol}, the fact that
clumps' masses evolve in time has direct implications for the amount of
magnetic support that the clump may have against its self-gravity.

\subsection{Production of Density Fluctuations. The Magnetic Case.}
\label{sec:dens_fluc_mag} 

In the magnetized case, the problem of density fluctuation production
becomes more complex, as the turbulent velocity field also produces
magnetic field fluctuations. In this section, we discuss two problems of
interest in relation to SF: The correlation of the density and magnetic
fluctuations, and the effect of the magnetic field on the PDF of density
fluctuations.

\subsubsection{Density-Magnetic Field Correlation} \label{sec:B-n_corr}

This is a highly relevant issue in relation to SF, as the
``standard'' model of mag\-net\-ic\-al\-ly-regulated SF 
\citep[hereafter SMSF; see, e.g., the reviews by][]{SAL87, Mousch91}
predicted that 
magnetic fields should provide support for the density fluctuations
(``clumps'') against their self-gravity, preventing collapse, except for
the material that, through AD, managed to lose its support (see the
discussion in \S \ref{sec:molec_turb_magn}). Thus, the strength of the
magnetic field induced in the turbulent density fluctuations is an
important quantity to determine.

Under perfect field-freezing conditions, the simplest scenario of a
fixed-mass clump threaded by an initially uniform magnetic field, and
undergoing an isotropic gravitational contraction implies that the field
should scale as $B \propto \rho^{2/3}$ \citep{Mestel66}, since the
density scales as $\rho \propto R^{-3}$, where $R$ is the radius of the
clump, while the flux-freezing condition implies that $B \propto
R^{-2}$. The assumption that the clump is instead oblate, or disk-like,
with the magnetic field providing support in the radial direction and
thermal pressure providing support in the direction perpendicular to its
plane, gives the scaling $B \propto \rho^{-1/2}$ \citep{Mousch76, Mousch91}

In a turbulent flow, however, the situation becomes more complicated.
``Clumps'' are not fixed-mass entities, but rather part of a continuum
that possesses random, chaotic motions. In principle, unless the
magnetic energy is {\it much} larger than the turbulent kinetic energy,
the compressive motions that form a clump can have any orientation with
respect to the local magnetic field lines, and thus the resulting
density enhancement may or may not be accompanied by a corresponding
magnetic field enhancement (cf.\ \S \ref{sec:eqs}). In particular,
\citet[][hereafter, 
PV03]{PV03} studied this problem analytically in the isothermal case, by
decomposing the flow into nonlinear, so-called ``simple'' waves
\citep[e.g.,][] {LL59}, which
are the nonlinear extensions of the well known linear MHD waves
\citep[e.g.,][] {Shu92}, having
the same three well-known modes: fast, slow, and Alfv\'en
\citep{Mann95}. For illustrative purposes, note that compressions along
the magnetic field lines are one instance of the {\it slow} mode, while
compressions perpendicular to the field lines (i.e., {\it magnetosonic}
waves) are an instance of the {\it fast} mode. As is well known,
Alfv\'en waves are transverse waves propagating along field lines, and
carry no density enhancement.  However, they can exert pressure, and the
dependence of this pressure on the density has been investigated by
\citet{MZ95} and PV03.

For simplicity and insight, PV03 considered the so-called
``1+2/3-dimensional'' case, also known as ``slab geometry'', meaning
that all three components of vector quantities are considered, but
their variation is studied with respect to only one spatial
dimension. For the Alfv\'en waves, they performed a linear perturbation
analysis of a circularly polarized wave. 
They concluded that each of the modes is characterized by a
different scaling between the magnetic pressure ($\propto B^2$) and the
density, as follows:
\barr
B^2 &\propto c_1 - \beta \rho  & \hbox{~~~~~slow,} \label{eq:slow_B_rho}\\
B^2 &\propto \rho^{2}       & \hbox{~~~~~fast,} \label{eq:fast_B_rho}\\
B^2 &\propto \rho^{\gamam}  & \hbox{~~~~~Alfv\'en,} \label{eq:Alfv_B_rho}
\earr
where $c_1$ is a constant, and $\gamam$ is a parameter that can take
values in the range (1/2,2) depending on the Alfv\'enic Mach number
\citep[see also][]{MZ95}. Note that eq.\ (\ref{eq:slow_B_rho}) implies
that for $\rho > c_1/\beta$ the slow mode disappears \citep{Mann95}, so
that only the fast and Alfv\'en modes remain. Conversely, note that, at
low density, the magnetic pressure due to the fast and Alfv\'en modes
becomes negligible in comparison with that due to the slow mode, which
approaches a constant. This implies that a log-log plot of $B$ vs.\
$\rho$ will exhibit an essentially constant value of $B$ at very small
values of the density. In other words, {\it at
low values of the density, the domination of the slow mode implies that
the magnetic field exhibits essentially no correlation with the density.}

\begin{figure}
\begin{center}
\includegraphics[scale=.45]{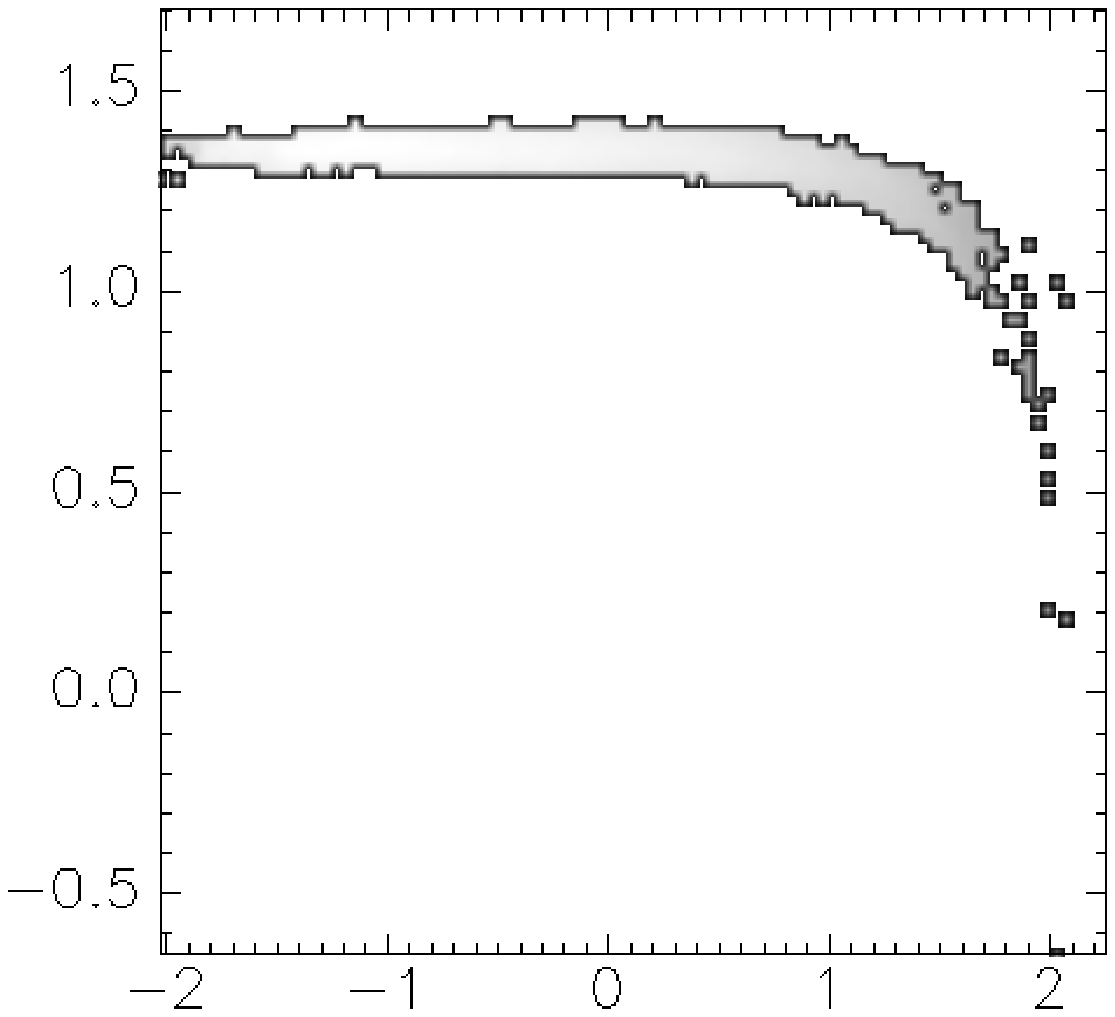}
\includegraphics[scale=.45]{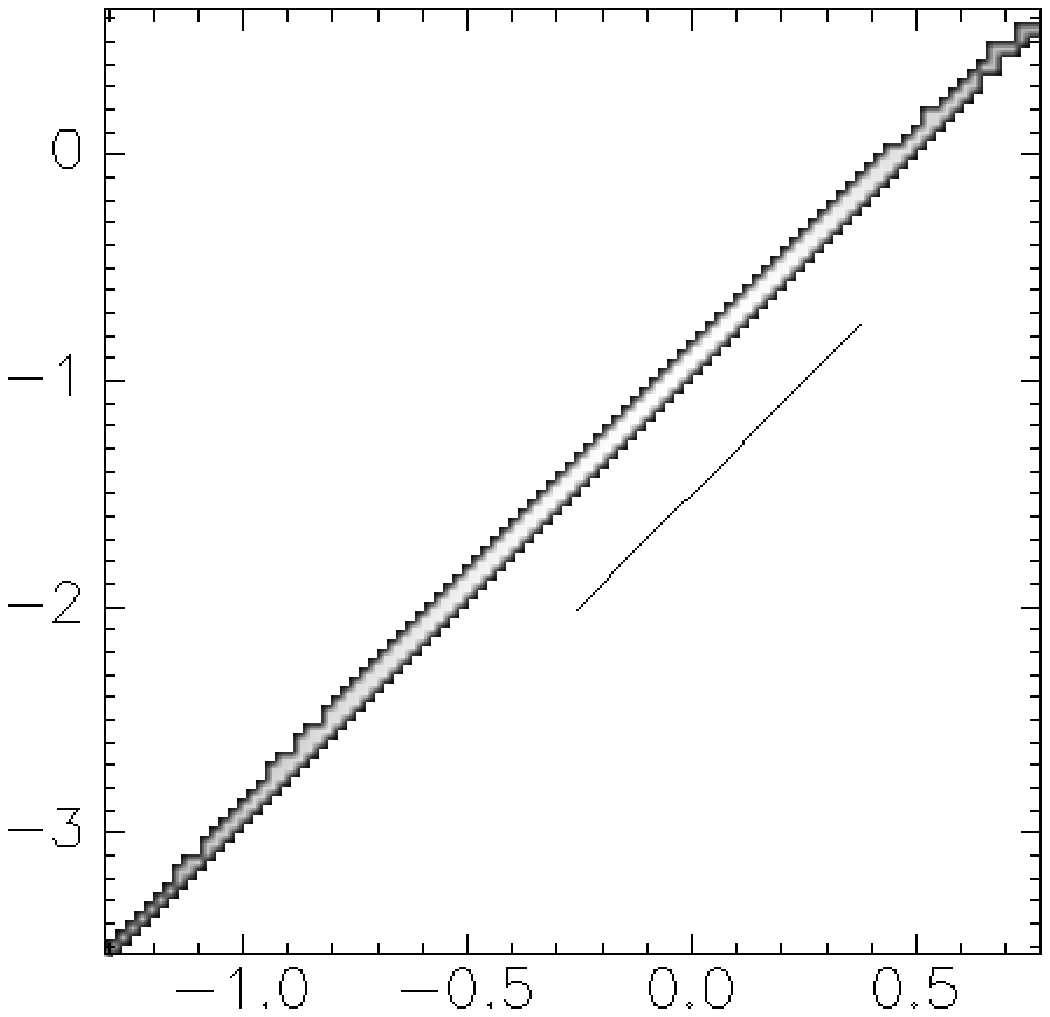}
\end{center}
\caption{Two-dimensional histograms of the grid cells in numerical
simulations in the $\ln B^2$ (vertical axis)-$\ln \rho$(horizontal axis)
space. The gray scale indicates the density of points in this space. {\it
Left:} A slab-geometry numerical simulation by PV03 dominated by the
slow mode, exhibiting the behavior indicated by eq.\
(\ref{eq:slow_B_rho}). {\it Right:} Same as the {\it left} panel, but
for a numerical simulation dominated by the fast mode, exhibiting the
behavior indicated by eq.\ (\ref{eq:fast_B_rho}).}
\label{fig:PV03_f1-2}
\end{figure}

PV03 were able to test these results numerically by taking
advantage of the slab geometry, which allowed to set up waves
propagating at well-defined angles with respect to the mean magnetic
field, and therefore being able to isolate, or nearly isolate, the three
different wave modes. The {\it left} panel of Fig.\ \ref{fig:PV03_f1-2} shows
the distribution of points in the $B^2$-$s$ space for a simulation
dominated by the slow mode, exhibiting the behavior outlined above,
corresponding to eq.\ (\ref{eq:slow_B_rho}). In
contrast, the {\it right} panel of Fig.\ \ref{fig:PV03_f1-2} shows the
distribution of points in the same space for a simulation dominated by
the fast mode, exhibiting the behavior indicated by eq.\
(\ref{eq:fast_B_rho}). 
\begin{figure}
\begin{center}
\includegraphics[scale=.35]{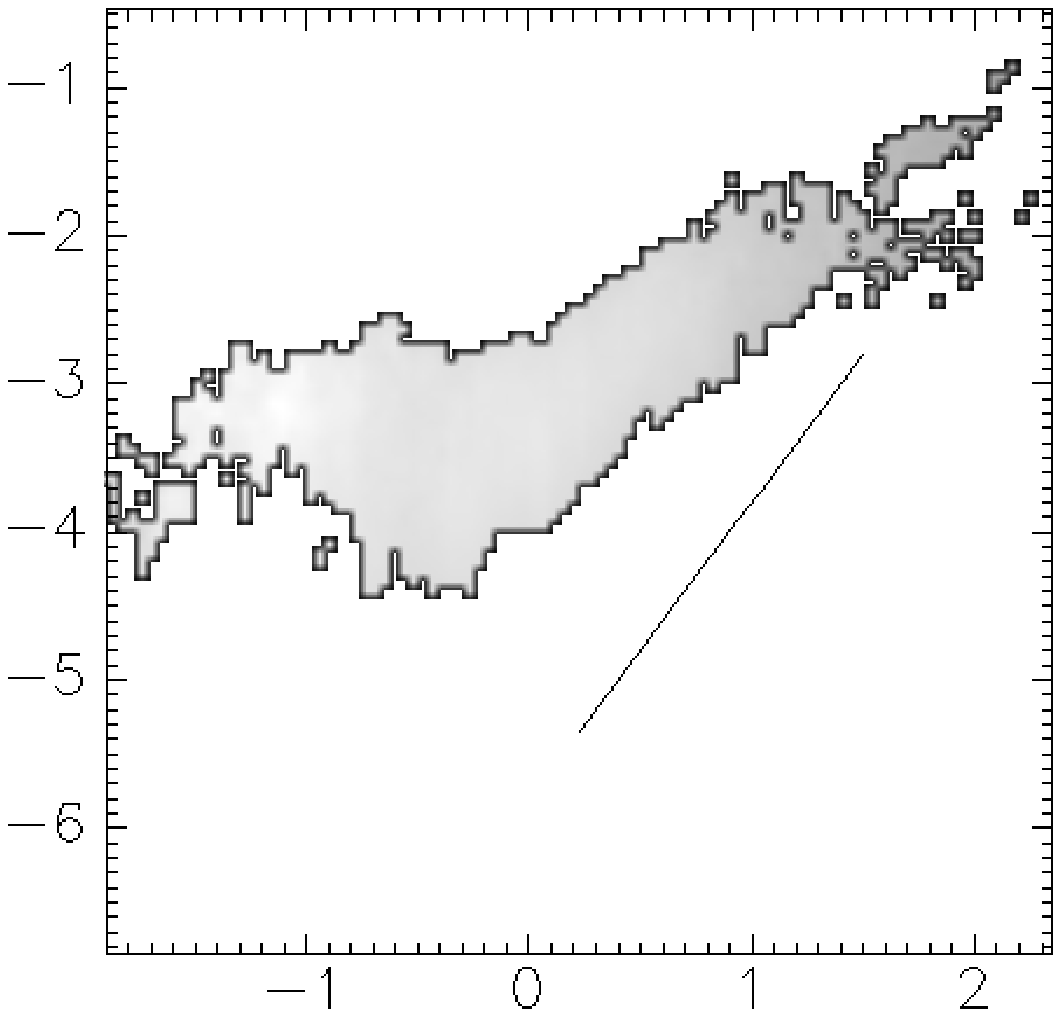}
\includegraphics[scale=.3]{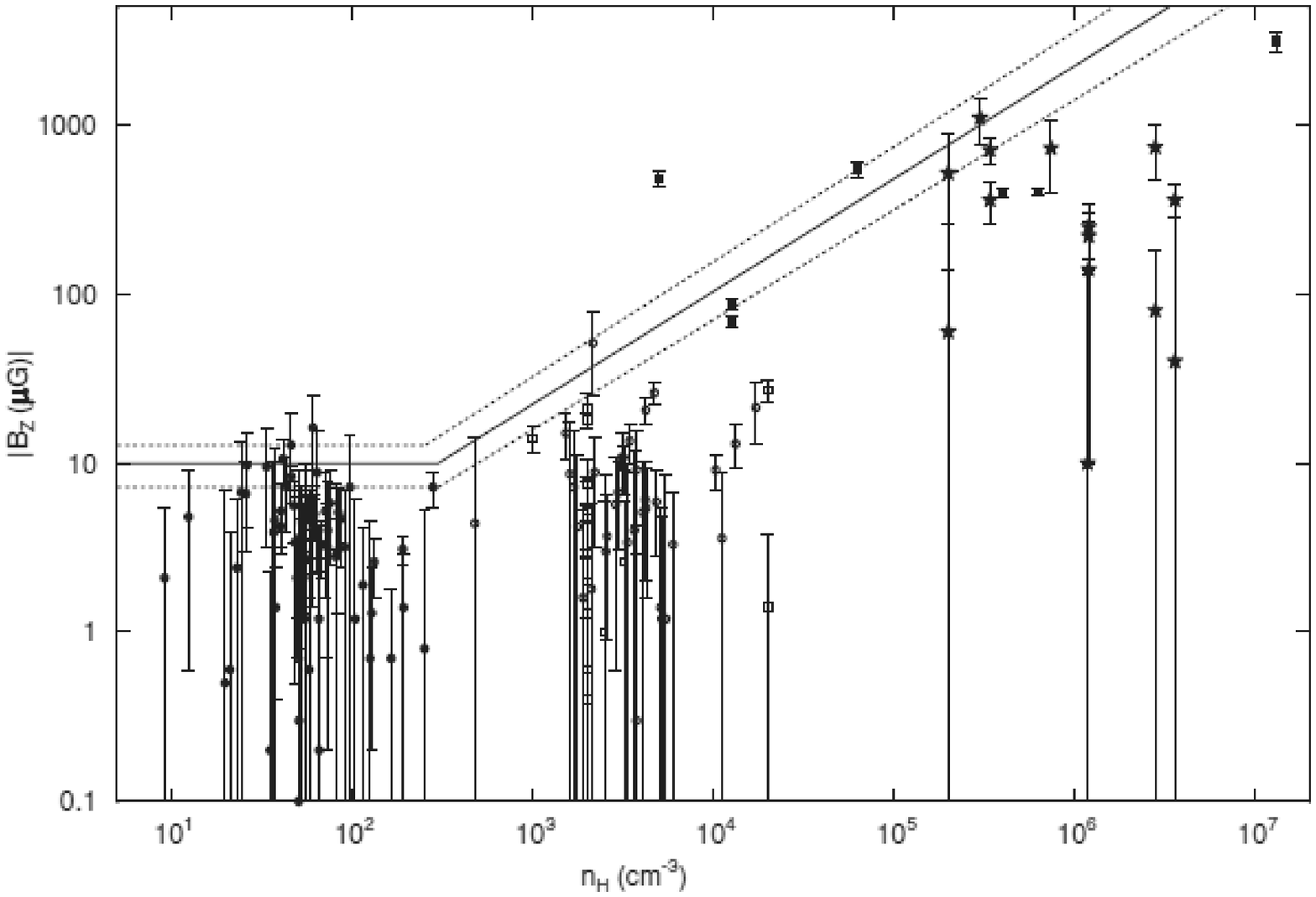}
\end{center}
\caption{{\it Left:} Two-dimensional histogram of the grid cells in the
$B^2$-$n$ space from a numerical simulation by PV03 in which both the
slow and the fast modes are active. At low densities, the slow mode
causes a density-independent magnetic field strength, while at higher
densities, the fast mode produces a positive correlation. The
straight-line segment has a slope of 2. {\it Right:}
Magnetic field strength determinations by Zeeman splitting observations
in molecular clouds, as compiled by \citet{Crutcher+10}. The rising
straight line segment has a slope $\approx 0.65$, implying $B^2 \propto
n^{1.3}$, in qualitative agreement with the numerical result.}
\label{fig:PV03_Crutcher+10}
\end{figure}

The most important conclusion from eqs.\
(\ref{eq:slow_B_rho})--(\ref{eq:Alfv_B_rho}) is that Now, in a turbulent
flow in which all modes are active, the net, average scaling of the
magnetic field with the density will arise from the combined effect of
the various modes. Moreover, since at low densities the values of $B$
produced by the fast and Alfv\'en modes are also small, while the field
strengths produced by the slow mode remain roughly constant, the field
fluctuations will be dominated by the latter mode at low densities, and
a roughly density-independent field strength is expected. Conversely, at
high densities, the slow mode disappears, while the contribution from
the fast and Alfv\'en modes will dominate, producing a field strength
that increases with increasing density.  Finally, because each mode
produces a different dependence of the magnetic field strength with the
density, we expect that the instantaneous value of the density at a
certain location in physical space is not enough to determine the value
of the magnetic field strength there. Instead, this value depends on the
{\it history of modes} of the nonlinear waves that have passed through
that location, naturally implying that, within a large cloud, a large
scatter in the measured values of the magnetic field is expected. The
expected net scaling of the field strength with the density is
illustrated in the {\it left} panel of Fig.\
\ref{fig:PV03_Crutcher+10}. These results are in qualitative agreement
with detailed statistical analyses of the magnetic field distribution in
the ISM \citep{Crutcher+10}, as illustrated in the {\it right} panel of
Fig.\ \ref{fig:PV03_Crutcher+10}.

\subsubsection{Effect of the Magnetic Field on the Density PDF}
\label{sec:magn_on_PDF}

According to the discussion in \S \ref{sec:dens_PDF}, the
dependence of the pressure on density determines the shape of the
density PDF, being a lognormal for the isothermal case, $\gamef =
1$. In the presence of a magnetic field, it would be natural to expect
that the magnetic pressure, which in general does not need to behave as an
isothermal polytrope, might cause deviations from the lognormal density
PDF associated to isothermal turbulent flows. 

However, the discussion above on the density-magnetic field correlation
(or rather, lack thereof), implies that the magnetic pressure does not
have a systematic effect on density fluctuations of a given amplitude,
as the value of the magnetic field is not uniquely determined by the
local value of the density. PV03 concluded that the effect of the
magnetic pressure was more akin to a random forcing in the turbulent
flow than to a systematic pressure gradient that opposes compression. As a
consequence, the underlying density PDF determined by the functional
form of the thermal pressure did not appear to be significantly affected
by the presence of the magnetic field, except under very special
geometrical setups in slab geometry, and that in fact are unlikely to
persist in a more general three-dimensional setup. The persistence of
the underlying PDF dictated by the thermodynamics in the presence of the
magnetic field is in agreement with numerical studies of isothermal MHD
turbulent flows that indeed have found approximately lognormal PDFs in
isothermal flows \citep[e.g.,][]{PN99, Ostriker+99, Ostriker+01, VG01,
Beresnyak+05}, and bimodal density PDFs in thermally-bistable flows
\citep{Gazol+09}, which we discuss in \S \ref{sec:pres_distr}.

\subsubsection{Evolution of the Mass-to-Magnetic Flux Ratio}
\label{sec:MFR_evol} 

The discussion in \S \ref{sec:clump_nature} implies that the mass
deposited in a clump by turbulent compressions is a somewhat ill-defined
quantity, depending on where and how one chooses to define the
``boundaries'' of the clump, and on the fact that its mass is
time-dependent. This has important implications for the so-called
mass-to-flux ratio (MFR) of the clump, and therefore, for the ability of
the magnetic field to support the clump against its self-gravity.

As is well known \citep{MS56}, a virial balance analysis implies that,
for a cloud of mass $M$ threaded by a uniform field $B$, gravitational
collapse can only occur if its MFR satisfies
\beq
\frac{M}{\Phi} > \left(\frac{M}{\Phi}\right)_{\rm crit} \equiv \left(\alpha
\pi^2 G\right)^{-1/2}, 
\label{eq:MFR_crit}
\eeq
where $\alpha$ is a constant of order unity whose precise value depends
on the shape and mass distribution in the cloud \citep[see,
e.g.,][]{MS56, NN78, Shu92}. Otherwise, the cloud is absolutely
supported by the magnetic field, meaning that the support holds
irrespective of the density of the cloud. In what follows, we
shall denote the MFR, normalized to this critical value, by $\lambda$.
Regions with $\lambda > 1$ are called {\it magnetically supercritical},
while those with $\lambda < 1$ are termed {\it magnetically subcritical}.
Traditionally, it has been assumed that the mass of the cloud is well
defined. However, our discussions above (\S \ref{sec:clump_nature})
suggest that it may be convenient to revisit these notions.

When considering density enhancements (``cores'') formed by turbulent
compressions within a cloud of size $L$, it is convenient to assume that
the initial condition for the cloud is one with uniform density and
magnetic field, and that the turbulence produces local fluctuations in
the density and the magnetic field strength. A simple argument advanced
by \citet{VS+05a} then shows that the MFR of a core of size $\ell$ and
MFR $\lambl$, must be within the range
\beq
\lambo \frac{\ell}{L} < \lambl < \lambo,
\label{eq:MFR_range}
\eeq
where $\lambo$ is the MFR of the whole cloud. The lower limit applies for
the case when the ``core'' is actually simply a subregion of the whole
cloud of size $L$, with the same density and magnetic field
strength. Since the density and field strength are the same, the mass of
the core simply scales as $(\ell/L)^3$, while the magnetic flux scales as
$(\ell/L)^2$.  Therefore, the MFR of a subregion of size $\ell$ scales
as $(\ell/L)$. Of course, this lower-limit extreme, corresponding to the
case of a ``core'' of the same density and field strength as the whole
cloud, is an idealization, since observationally such a structure cannot
be distinguished from its parent cloud. Nevertheless, as soon as some
compression has taken place, the core will be observationally
distinguishable from the cloud (for example, by using a tracer that is
only excited at the core's density), and the measurement of the MFR {\it
in the core} will be bounded from below by this limit.

On the other hand, the upper limit corresponds simply to the case where
the entire cloud of size $L$ has been compressed isotropically to a size
$\ell$, since in this case both the mass and the magnetic flux are
conserved, and so is the MFR. 

This reasoning has the implication that the MFR that is {\it measured}
in a core within a cloud must be smaller than that measured for the
whole cloud, as long as the condition of flux-freezing holds.  Note that
one could argue that this is only an observational artifact, and that
the physically relevant mass is that associated to the whole flux tube
the core belongs to, but this is only reflecting the ambiguity discussed
above concerning the masses of clumps. In practice, {\it the physically
relevant mass for the computation of the MFR is the one responsible for
the local gravitational potential well against which the magnetic field
is providing the support}, and this mass is precisely the mass of the
core, not the mass along the entire flux tube, especially when phase
transitions are involved.

\begin{figure}
\begin{center}
\includegraphics[scale=.27]{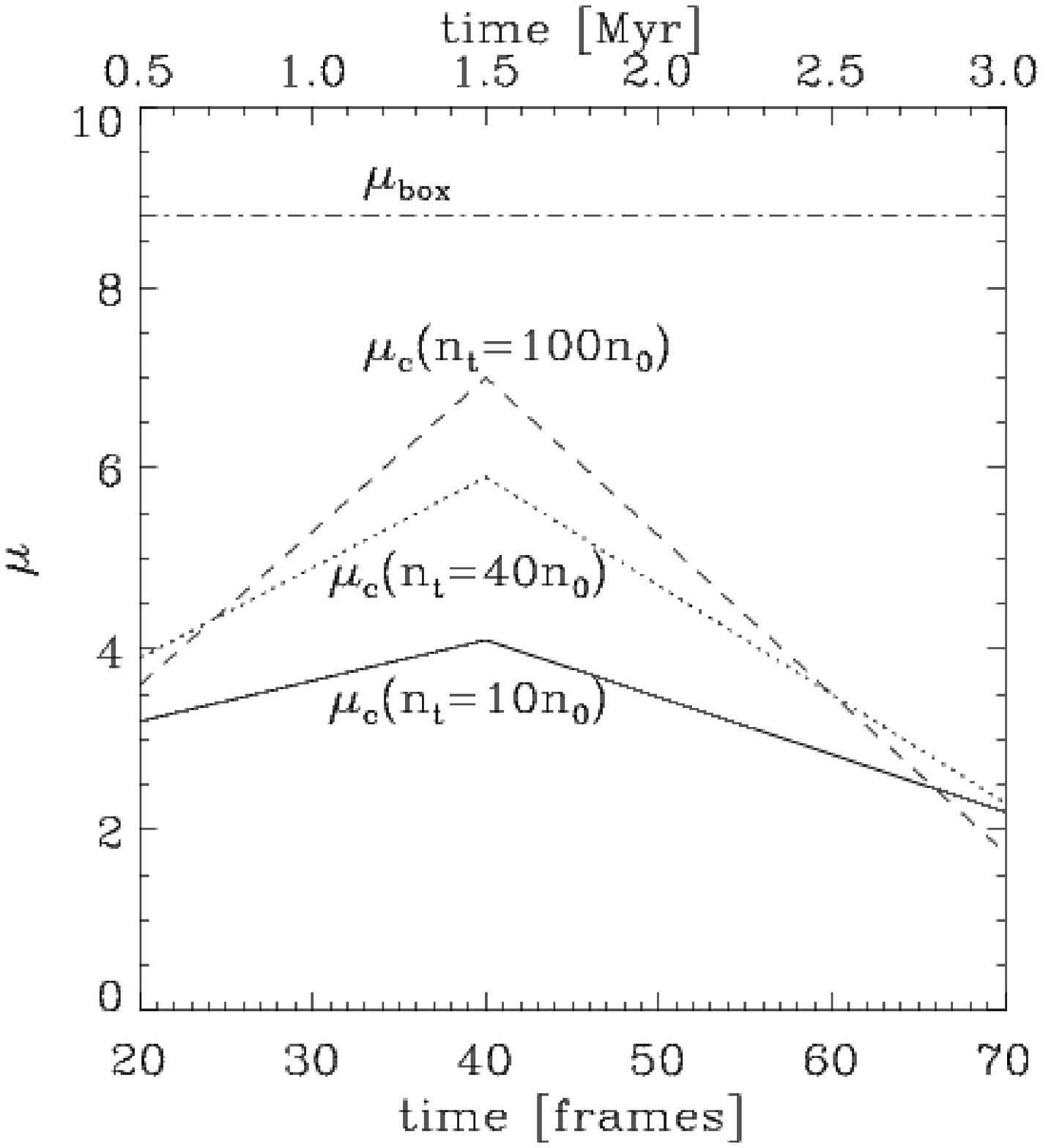}
\includegraphics[scale=.27]{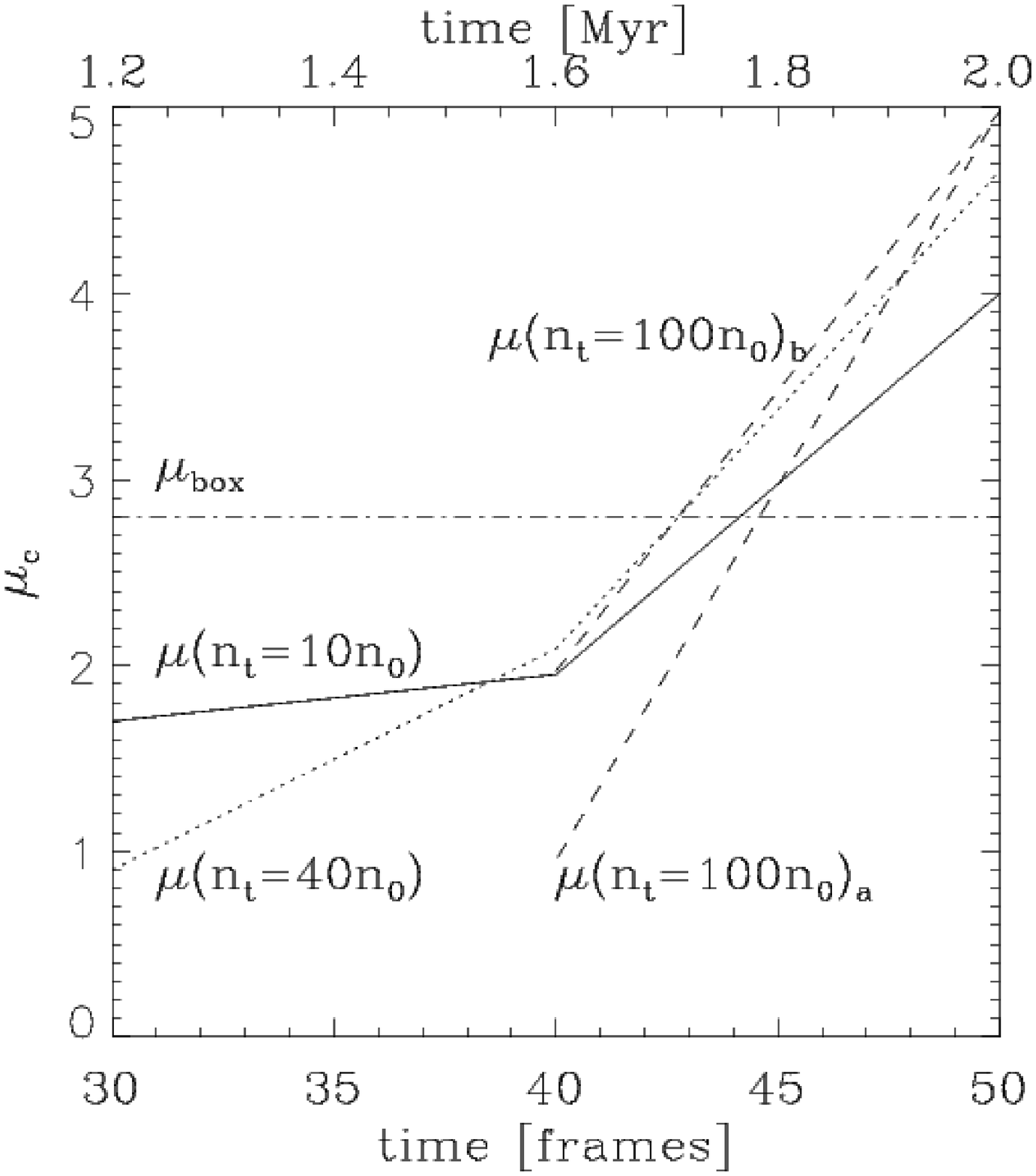}
\end{center}
\caption{Evolution of the mass-to-flux ratio (MFR, denoted $\mu$ in
these plots),
normalized to the critical value, in cores formed in numerical
simulations of continuously driven MHD isothermal turbulence by
\citet{VS+05a}. The cores are defined as connected sets of grid cells
with density above a density threshold $\nt$, and are followed over
time. Three values of $\nt$ are used, illustrating how the value of the
MFR depends and evolves as successively more internal regions of the
density fluctuation are considered. {\it Left:} Evolution of $\lambda$
in a core that does not collapse. At all values of $\nt$, the MFR first
increases and then decreases again. {\it Right:} A core that does
collapse. In this panel, two lines are shown at the largest value of
$\nt$, because the parent clump splits into two cores at this
threshold. In both panels, the rate of variation of $\lambda$ is larger
for larger values of $\nt$, and $\lambda$ for the innermost region ($\nt
= 100 n_0$, where $n_0$ is the mean density of the simulation) is seen
to start {\it lower} than that of the envelope, and to overtake it as
the degree of mass concentration is increased, in this case as a
consequence of numerical diffusion, which plays a role analogous to that
of AD.}
\label{fig:VS05_f7-8}
\end{figure}

Note also that, as discussed in \S \ref{sec:clump_nature}, the mass of
the clump must be evolving in time. If the compression is occurring mostly
along field lines (since in this direction the magnetic field presents
no resistance to it), then the magnetic flux remains roughly constant,
while the mass increases at first, and later it possibly decreases if
the clump begins to re-expand. Otherwise, if the core becomes massive
enough that it becomes gravitationally unstable {\it and} supercritical,
it must begin to collapse gravitationally (Fig.\
\ref{fig:VS05_f7-8}). At this point, the rapid density enhancement at
the core in turn enhances the action of AD \citep[e.g.,][see also \S
\ref{sec:eqs}]{SAL87, Mousch91}, causing the magnetic flux to escape the
core,\footnote{Note that this ``escape'' is meant in a Lagrangian sense,
i.e., following the flow. That is, considering a certain fluid parcel as
it contracts, AD causes the flux to be ``left behind'' from the fluid
particles that make up the parcel. Conversely, in an Eulerian sense, the
magnetic flux remains fixed, but the fluid parcel increases its mass in
this frame.} so that the latter eventually acquires a larger value of
$\lambda$ than its envelope. Thus, the prediction from this dynamic scenario
of core formation is that cores in early stages of evolution should
exhibit smaller values of the MFR than their envelopes, while cores at
more advanced stages should exhibit larger values of the MFR than their
envelopes. Evidence in this direction has begun to be collected
observationally \citep{Crutcher+09}, as well as through synthetic
observations of numerical simulations \citep{Lunttila+09}.

Before closing this section, an important remark is in order. Recent
numerical and observational evidence (cf.\ Sec \ref{sec:molec_turb})
suggests that the ``turbulence'' in star-forming molecular clouds may
actually consist of a hierarchy of gravitational contraction motions,
rather than of random, isotropic turbulence. In such a case, the
physical processes discussed in this section are still applicable,
noting that the converging flows that produce the clumps may be driven
by larger-scale gravitational collapse rather than by random turbulent
compressions, and the only part of the previous discussion that ceases
to be applicable is the possibility that some cores may fail to collapse
and instead re-expand. If the motions all have a gravitational origin,
then essentially all cores must be on their way to collapse. It is
worth pointing out that in this case, what drives the collapse of an
apparently subcritical core is the collapse of its parent, supercritical
structure.

\section{Turbulence in the Multiphase ISM} \label{sec:turb_therm}

In the previous sections we have separately discussed two different
kinds of physical processes operating in the ISM: radiative heating and
cooling, and compressible MHD turbulence in the special case of
isothermality. However, since both operate simultaneously in the atomic
ISM, it is important to understand how they interact with each other,
especially because the mean density of the Galactic ISM at the Solar
galactocentric radius, $\langle n \rangle \sim 1 \pcc$, falls precisely
in the thermally unstable range. This problem has been investigated
numerically by various groups \citep[e.g.,][]{HP99, HP00, WF00, KI00,
KI02, VS+00a, VS+03, VS+06, VS+07, VS+11, Hennebelle+08, Banerjee+09,
Gazol+01, Gazol+05, Gazol+09, KN02, SS+02, PO04, PO05, AH05, AH10,
Heitsch+05, HA07}, and in this section we review their main results.

\subsection{Density PDF in the Multiphase ISM}
\label{sec:pres_distr} 

\begin{figure}
\begin{center}
\includegraphics[scale=.29]{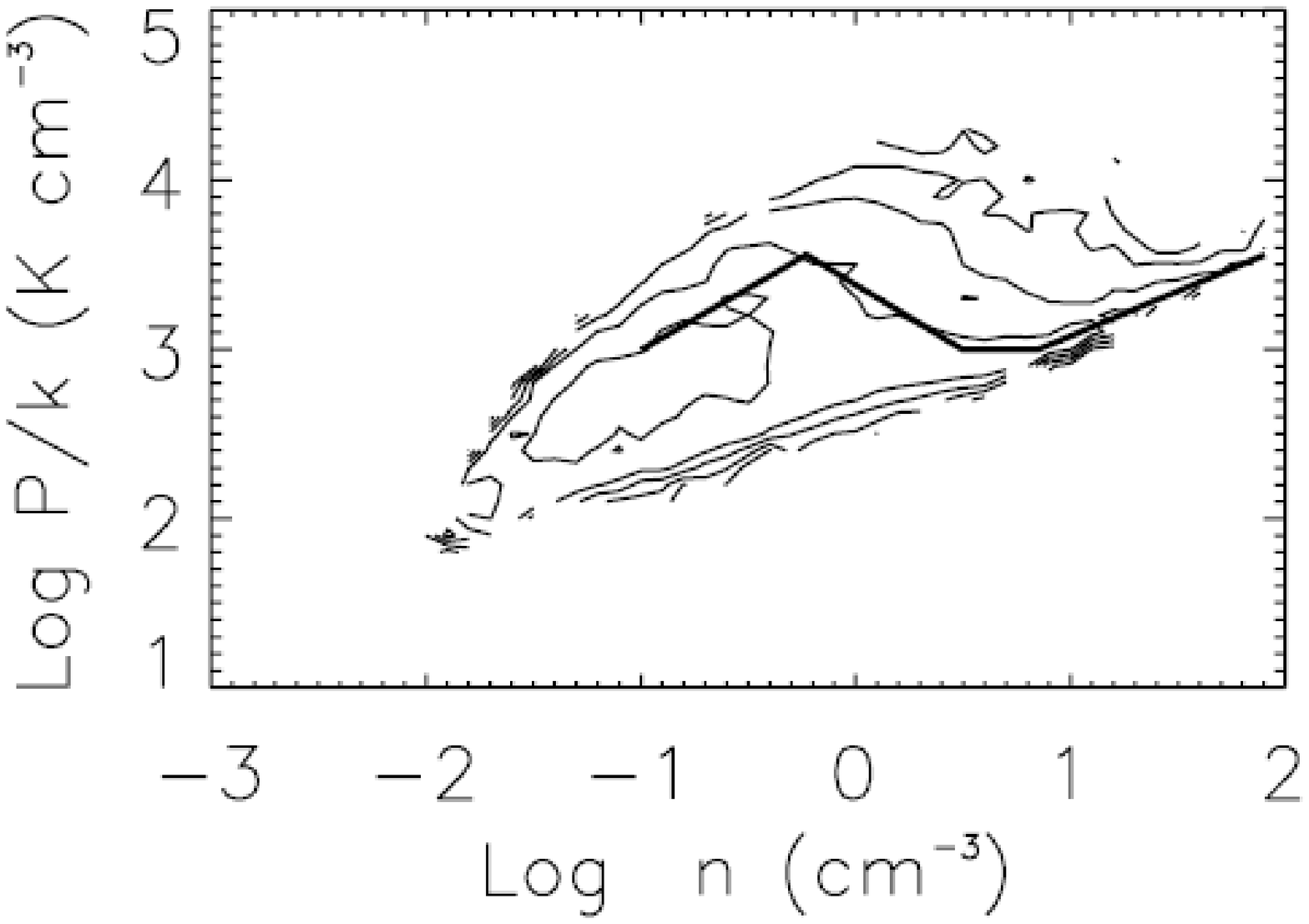}
\includegraphics[scale=.32]{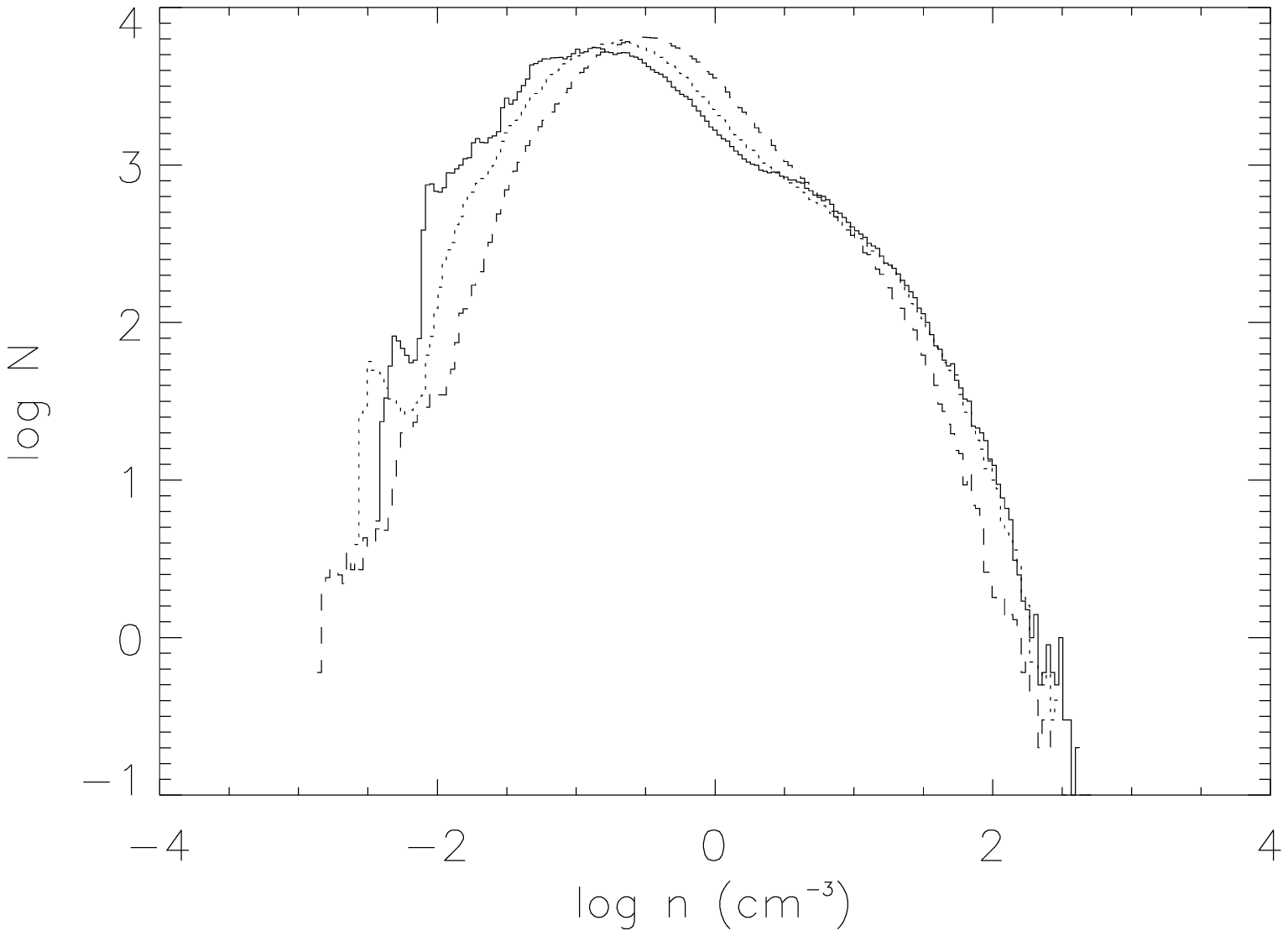}
\end{center}
\caption{{\it Left:} Two-dimensional histogram of the grid cells in the
pressure-density diagram for a two-dimensional simulation of turbulence
in the thermally-bistable atomic medium, with {\it rms} velocity
dispersion of $9 \kms$, a numerical box size of 100 pc, and the
turbulent driving applied at a scale of 50 pc. {\it Right:} Density PDF
in simulations like the one on the {\it left} panel, but with three different
values of the {\it rms} velocity: $4.5 \kms$ ({\it solid line}), $9
\kms$ ({\it dotted line}), and $11.3 \kms$ ({\it dashed line}). The
bimodality of the PDF is seen to to become less pronounced as the rms
velocity increases, with a single power-law tail developing in the
density range between the values where the peaks would be
otherwise located. From \citet{Gazol+05}.  }
\label{fig:P_rho_pdfs_TI}
\end{figure}

A key parameter controlling the interaction between turbulence and
net cooling is the ratio $\eta \equiv \tcool/\tturb$, where $\tcool
\sim k T/(n \Lambda)$ is the cooling time, with $k$ being the Boltzmann
constant, and $\tturb \sim L/U$ is the turbulent crossing time. The
remaining symbols have been defined above. In the limit $\eta
\gg 1$, the dynamical evolution of the turbulent compreesions occurs much more
rapidly than they can cool, and therefore the compressions behave nearly
adiabatically. Conversely, in the limit $\eta \ll 1$, the fluctuations
cool down essentially instantaneously while the turbulent compression is
evolving, and thus they tend to reach the thermal equilibrium pressure
$\Peq$ as soon as they are produced\footnote{It is often believed that
fast cooling directly implies isothermality.  However, this is a
misconception. While it is true that fast cooling is a necessary
condition for approximately isothermal behavior, the reverse implication
does not hold. Fast cooling only implies an approach to the thermal
equilibrium condition, but this need not be isothermal. The precise form
of the effective equation of state depends on the details of the
functional dependence of the heating and cooling functions on the
density and temperature.} \citep{Elm91, PVP95, SS+02, VS+03,
Gazol+05}. Because in a turbulent flow velocity fluctuations of a wide
range of amplitudes and size scales are present, the resulting density
fluctuations in general span the whole range between those limits, and
the actual thermal pressure of a fluid parcel is not uniquely determined
by its density, but rather depends on the details of the velocity
fluctuation that produced it. This causes a scatter in the values of the
pressure around the thermal-equilibrium value in the pressure-density
diagram (Fig.\ \ref{fig:P_rho_pdfs_TI}, {\it left} panel), and also
produces significant amounts of gas (up to nearly half of the total
mass) with densities and temperatures in the classically forbidden
thermally unstable range \citep{Gazol+01, dAB05, AH05, ML+05}, a result
that has been encountered by various observational studies as well
\citep[e.g.,][]{Dickey+78, Heiles01}. In any case,  the
tendency of the gas to settle in the stable 
phases still shows up as a multimodality of the density PDF, which
becomes less pronounced as the {\it rms} turbulent velocity increases
(Fig.\ \ref{fig:P_rho_pdfs_TI}, {\it right} panel).

\subsection{The Formation of Dense, Cold Clouds and Clumps} \label{sec:cloud_form}

\subsubsection{The Non-Magnetic Case}
\label{sec:cloud_form_hd} 

\begin{figure}
\begin{center}
\includegraphics[scale=.38]{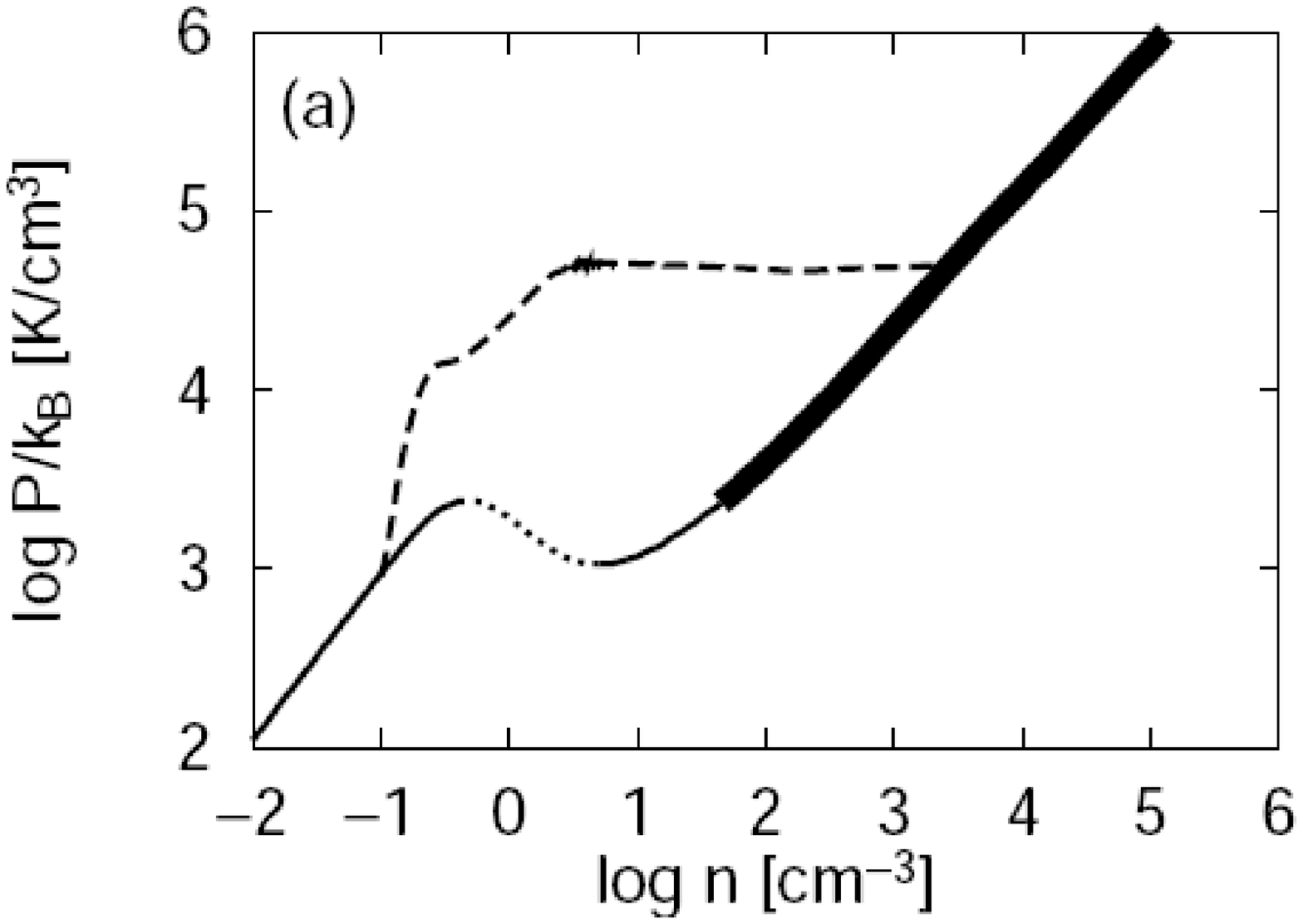}
\includegraphics[scale=.51]{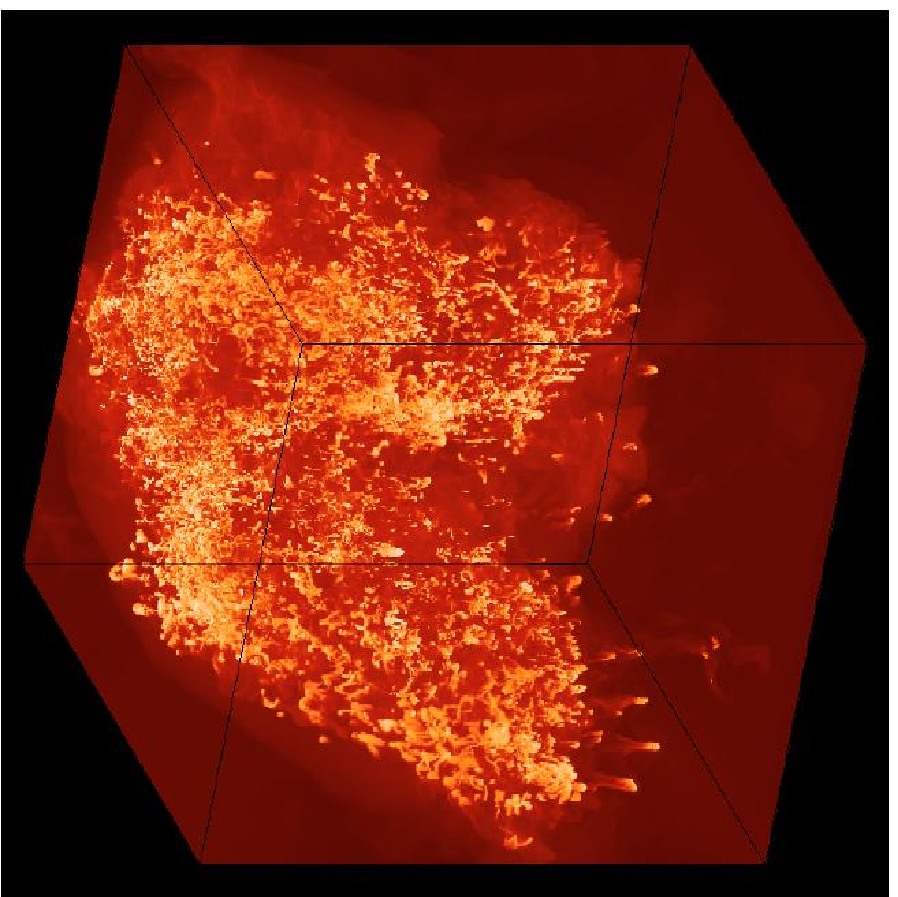}
\end{center}
\caption{{\it Left:} Evolutionary path ({\it dashed line}) in the $P$
vs.\ $\rho$ diagram of a fluid parcel initially in the WNM after
suffering a transonic compression that nonlinearly triggers TI. The {\it
solid} and {\it dotted lines} show the locus of $\Peq(\rho)$, the solid
sections corresponding to linear stability and the dotted ones to linear
instability. The solid section to the left of the dotted line
corresponds to the WNM and the one at the right, to the CNM. The
perturbed parcel evolves from left to right along the dashed line. From
\citet{KI00}. {\it Right:} Projected (or {\it column}) density structure
of the resulting GMC in a numerical simulation of its formation by
colliding WNM streams, and its subsequent evolution. The numerical box
has a size of 15 pc on a side, and the resolution is $1200^3$ grid
cells. The ``GMC'' is seen to consist of the agglomeration of a huge
number of small clumps, which have formed by fragmentation caused by the
action of combined instabilities in the compressed gas. From
\citet{AH10}. }
\label{fig:nonlin_condens}
\end{figure}

A very important consequence of the interaction of turbulence (or, more
generally, large-scale coherent motions of any kind) and TI is that the
former may {\it nonlinearly} induce the latter. Indeed, \citet[][see
also Koyama \& Inutsuka 2000]{HP99} showed that transonic (i.e., with
$\Ms \sim 1$) compressions in the WNM can compress the medium and bring
it sufficiently far from thermal equilibrium that it can then undergo a
phase transition to the CNM (Fig.\ \ref{fig:nonlin_condens}, {\it left}
panel). This process amounts then to producing a cloud with a density
up to $100\times$ larger than that of the WNM by means of only moderate,
transonic compressions. This is in stark contrast
with the process of producing density fluctuations by pure supersonic
compressions in, say, an isothermal medium, in which such density
contrasts would require Mach numbers $\Ms \sim 10$. It is worth noting
that the turbulent velocity dispersion of $\sim 8$--11$\kms$ in the warm
Galactic ISM \citep{KH87, HT03} is, precisely, transonic.

Moreover, the cold clouds formed by this mechanism have typical sizes
given by the size scale of the compressive wave in the transverse
direction to the compression, rather than having to be of the same size
scale as the fastest growing mode of TI, which is very small ($\lesssim 0.1$
pc; cf.\ \S \ref{sec:thermodynamics}). The initial stages of this
process may produce thin CNM sheets \citep{VS+06}, which are in fact
observed \citep{HT03}. However, such sheets are quickly destabilized,
apparently by a combination of nonlinear thin shell
\citep[NTSI;][]{Vishniac94}, Kelvin-Helmholtz and Rayleigh-Taylor
instabilities \citep{Heitsch+05}, fragmenting and becoming
turbulent. This causes the clouds to become a complex mixture of cold
and warm gas, where the cold gas is distributed in an intrincate network
of sheets, filaments and clumps, possibly permeated by a dilute, warm
background. An example of this kind of structure is shown in the {\it
right} panel of Fig.\ \ref{fig:nonlin_condens}.

A noteworthy feature of the clouds and clumps formed by TI is that,
contrary to the case of density fluctuations in single-phase media, they
can have more clearly defined and long-lasting boundaries. This is
because their boundaries may be defined by the locus of the interface
between the cold and warm phases, which, once formed, tends to persist
over long timescales compared to the dynamical time, because the two
phases are essentially at the same pressure. {\it Under
quasi-hydrostatic conditions}, these boundaries would have little or no
mass flux across them \citep[i.e., they are {\it contact
discontinuities}; see, e.g.,][]{Shu92}. Any mass exchange that managed
to happen would be due to evaporation or condensation, occurring when the
thermal pressure differs from the saturation value between the phases
\citep[e.g.,][]{ZP69, PB70, Nagashima+05, Inoue+06}.  The
latter two papers have in fact proposed that such evaporation may
contribute to the driving of interstellar turbulence, although the
characteristic velocities they obtained ($\lesssim 1 \kms$) appear to be
too small for this to be the dominant mechanism for driving the large
scale ISM turbulence, with characteristic speeds of $\sim 10 \kms$.

\begin{figure}
\begin{center}
\includegraphics[scale=.27]{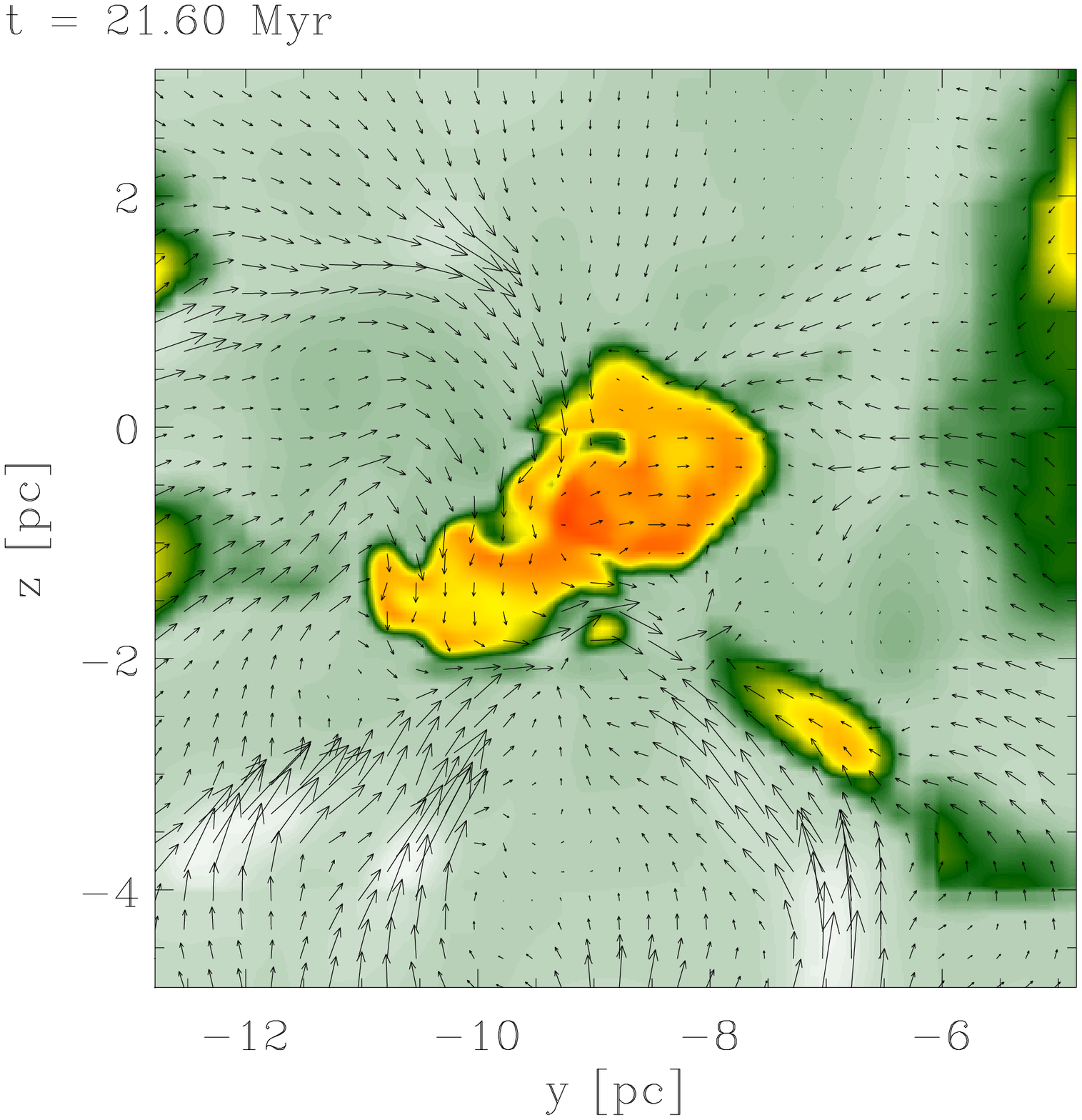}
\includegraphics[scale=.27]{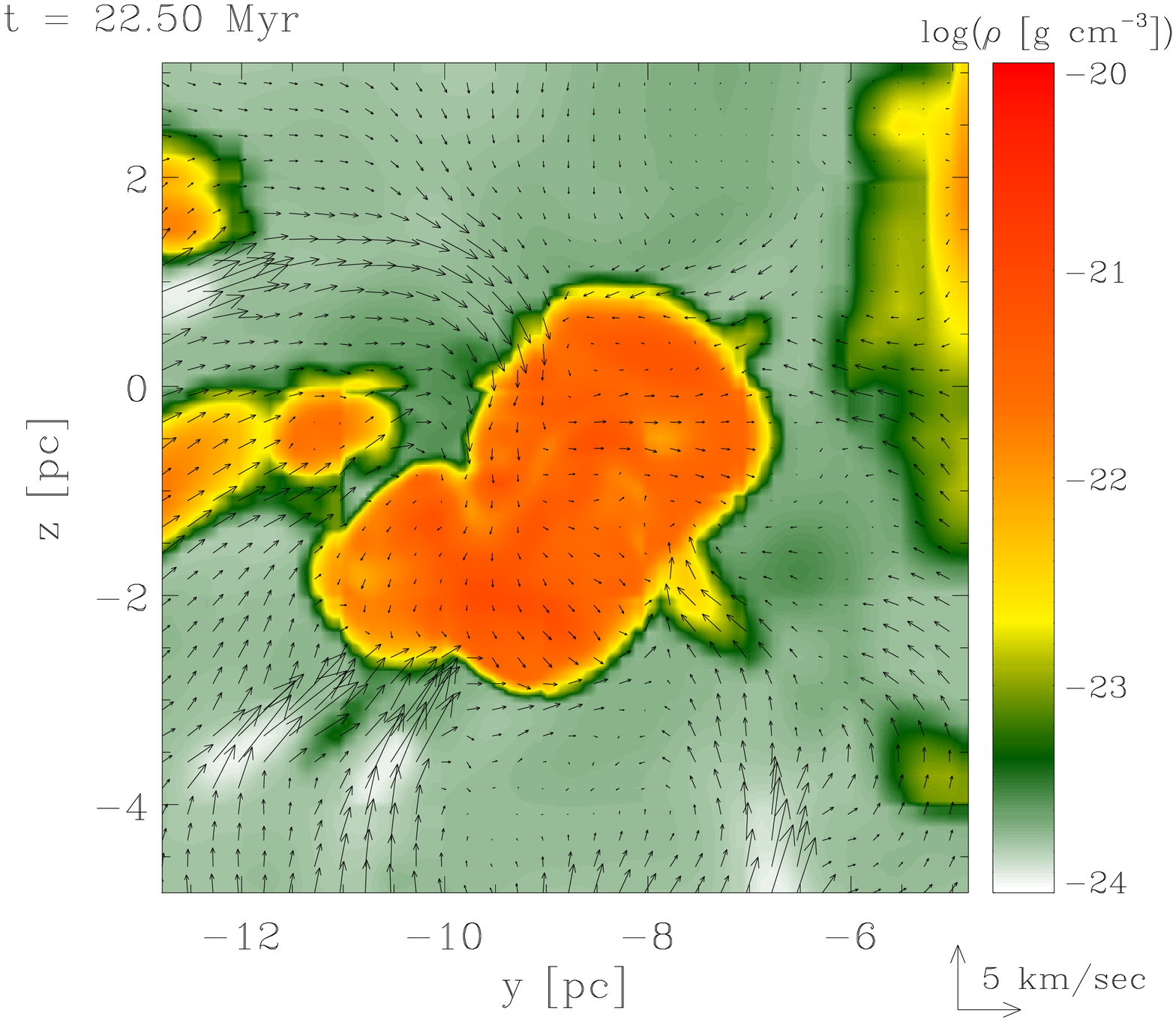}
\end{center}
\caption{Density cuts through the plane $x=2.5$ pc in an adaptive-mesh
refinement (AMR) numerical simulation of molecular cloud formation by
\citet{Banerjee+09}, illustrating the clump growth mechanism. The
numerical box size is 256 pc, and the maximum resolution is 0.06 pc. The
arrows show the projection of the velocity field on this plane. {\it
Left:} The clump at time $t=21.6$ Myr. {\it Right:} The clump at $t=
22.5$ Myr. Comparing the two times the growth of the clump is
evident. The velocity field is seen to generally point towards the
clump, indicating that material from the diffuse external medium is
entering the clump, causing its growth.}
\label{fig:clump_growth}
\end{figure}

However, in the presence of large-scale ($> 10$ pc) and large-amplitude
($\gtrsim 10$ km s$^{-1}$) motions, corresponding either to the
supernova-driven global ISM turbulence, to larger-scale instabilities,
such as the magneto-Jeans \citep[e.g.,][]{KO01} or magneto-rotational
\citep{BH91} ones, or simply to the passage of spiral arms, the
{\it nonlinear} triggering of TI implies that the fronts bounding the
clouds and clumps are not contact discontinuities, but rather {\it phase
transition fronts} -- structures analogous to shocks, with large
density, temperature and velocity jumps, but without the need for
locally supersonic velocities. Across such fronts, a substantial mass
flux occurs \citep{VS+06, Banerjee+09}. This is mechanism is illustrated
in Fig.\ \ref{fig:clump_growth}, which clearly shows the rapid growth of
a clump by accretion of diffuse material in a numerical simulation. This
is in stark contrast with early ideas that the clumps grew by
coagulation of tiny cloudlets on very long timescales \citep[$\sim 100$ Myr;
e.g.,][]{Kwan79}.
%

It should be emphasized, however, that the density does not necessarily
always present a sharp jump between the clump and its surroundings. As
discussed in \S \ref{sec:pres_distr}, the presence of turbulence in the
diffuse medium also implies a certain degree of mixing, and the
existence of a certain fraction of the mass that is traversing the
unstable range. In Fig.\ \ref{fig:clump_growth} this can be observed as
the greenish regions, especially near the right edge of the {\it left}
panel.

We conclude then that, although in thermally bistable flows clump
boundaries are in general better defined than in turbulent isothermal
flows because of the density jumps induced by the thermal bistability,
this does not imply that they are impenetrable boundaries that restrict
the flow of the medium. Rather, the clumps are formed and then grow by
accretion of diffuse material across these phase transition fronts.

Finally, it is important to note that, in the scenario of GMC formation
described above, the compressions in the WNM tend to initially form thin
sheets of CNM \citep{VS+06}, in agreement with observations of CNM
clouds \citep{HT03}. These sheets, however, grow by accretion of diffuse
gas, fragmenting and becoming turbulent due to the combined action of
various instabilities \citep{VS+06, Heitsch+06}, so that GMCs may
actually consist of huge conglomerates of small clumps, as illustrated
in the {\it right} panel of Fig.\ \ref{fig:nonlin_condens}. This is
consistent with the observed clumpy structure of GMCs \citep[][sec.\
VII]{Blitz93}.
%

\subsubsection{The Magnetic Case} \label{sec:cloud_form_mag}

In the presence of a magnetic field, the process of cloud formation by
phase transitions to the cold phase requires further considerations.
First, the orientation of the compressive motion relative to that of the
magnetic field strongly influences the ability of the compression to
trigger a transition to the dense phase. \citet{HP00} investigated this
problem by means of numerical simulations with slab (1+1/2D) geometry,
finding that, for a certain value of the magnetic field strength, and a
given sonic Mach number of the compression, there exists a maximal angle
between the direction of compression and the direction of the magnetic
field beyond which no phase transition is induced. They found this angle
to typically lie between 20 to 40 degrees, for typical values of the
warm neutral medium.

\citet{HP00} also found that, when the formation of a cloud
does occur, either the field is re-oriented along the compression (in
the case of weak fields), or the flow is re-oriented along field lines
(in the case of strongter fields), and the accumulation of gas to form
the clump ends up being aligned with the magnetic field.  In addition,
\citet{II08} have found that compressions perpendicular to the magnetic
field strongly inhibit the formation of dense, molecular-type clouds,
and that, in this case, only diffuse HI clouds manage to form. As a
consequence, the discussion of cloud formation can be made in terms of
compressions parallel to the magnetic field without loss of
generality. We will take up this problem again in \S
\ref{sec:molec_turb}, when we discuss the onset of gravitational
collapse of the clouds.

\section{The Nature of the Turbulence in the various ISM Components}
\label{sec:turb_multi} 

\subsection{Generalities} \label{sec:generalities}

As discussed in the previous sections, the ionized and atomic components
of the ISM consist of gas in a wide range of temperatures, from $T \sim
10^6$ K for the HIM, to $T \sim 40$ K for the CNM. In particular,
\citet{HT03} report temperatures in the range $500 < T < 10^4$ K for the
WNM, and in the range $10 < T < 200 $ K for the CNM. The WIM is expected
to have $T \sim 10^4$ K. This implies that the adiabatic
sound speed, given by \citep[e.g.,][]{LL59}
\beq 
\cs = \sqrt{\frac{\gamma k T}{\mum \mh}} \approx  10.4 \kms \left(\frac{T}
{10^4 {\rm K}} \right)^{1/2},
\label{eq:cs}
\eeq
will also exhibit large fluctuations in the medium. In the second
equality, we have used $\mum = 1.27$. In the following
sections we discuss the implications of these ranges for the various ISM
components.

\subsection{The Hot Ionized medium} \label{sec:turb_HIM}

At temperatures $T \sim 10^6$ K, the sound speed in the HIM is $\sim 100
\kms$, much larger than the velocity dispersion in the general
ISM, $\sim 10 \kms$. Thus, except in the immediate vicinity of supernova
explosions, where the velocities can reach thousands of$\kms$, the HIM
in general is expected to behave nearly incompressibly. Moreover,
because the density is very low ($\lesssim 10^{-2} \pcc$), the cooling
time ($\tcool \sim kT/n\Lambda$; cf.\ Sec.\ \ref{sec:pres_distr}) is very long
(a few tens of Myr), and the medium is then expected to behave roughly
adiabatically, at least up to scales of a few hundred parsecs.

\subsection{The Warm Ionized Medium} \label{sec:turb_WIM}

Collecting measurements of
interstellar scintillation (fluctuations in amplitude and phase of radio
waves caused by scattering in the ionized ISM) from a variety of
observations, \citet{Armstrong+95} estimated the power spectrum of
density fluctuations in the WIM, finding that it is consistent with a
\citet{K41} spectrum, a result expected for weakly compressible flows
\citep{Bayly+92}, on scales $10^8 \lesssim L \lesssim 10^{15}$ cm. 

More recently, using data from the Wisconsin H$\alpha$ Mapper
Observatory, \citet{CL10} have been able to extend the spectrum to
scales $\sim 10^{19}$ cm, suggesting that the WIM behaves as an
incompressible turbulent flow over size scales spanning more than 10
orders of magnitude. This suggestion is supported also by the results of
\citet{Hill+08} who, by measuring the distribution of H$\alpha$ emission
measures in the WIM, and comparing with numerical simulations of
turbulence at various Mach numbers, concluded that the sonic Mach number
of the WIM should be $\sim 1.4$--2.4. Although the WIM is ionized, and
thus should be strongly coupled to the magnetic field, the turbulence
then being magnetohydrodynamic (MHD), Kolmogorov scaling should still
apply, according to the theory of incompressible MHD fluctuations
\citep{GS95}. The likely sources of kinetic energy for these turbulent
motions are stellar energy sources such as supernova explosions
\citep[see, e.g.,][]{MK04}.

\subsection{The Atomic Medium} \label{sec:atomic_turb}

In contrast to the relatively simple and clear-cut situation for the
ionized ISM, the turbulence in the neutral (atomic and molecular) gas is
more complicated. According to the discussion in \S
\ref{sec:generalities}, the temperatures in the atomic gas may span a
continuous range from a few tens to several thousand
degrees. Additionally, \citet{HT03} report column density-weighted {\it
rms} velocity dispersions $\sigma_v \sim 11
\kms$ for the WNM, and typical internal motions of $\Ms \sim 3$ for the
CNM. It is thus clear that the WNM is transonic
($\Ms \sim 1$), while the CNM is moderately
supersonic.  This occurs because the atomic gas is thermally bistable,
and because transonic compressions in the WNM can nonlinearly induce TI
and thus a phase transition to the CNM (\S \ref{sec:pres_distr}). Thus, the
neutral atomic medium is expected to consist of a complex mixture of gas
spanning over two orders of magnitude in density and temperature.

It is worth noting that early pressure-equilibrium models
\citep[e.g.,][]{FGH69, MO77} proposed that the unstable phases were
virtually nonexistent in the ISM, but the observational and numerical
results discussed in \S \ref{sec:pres_distr} suggest that a significant
fraction of the atomic gas mass lies in the unstable range, 
transiting between the stable phases. Also, numerical simulations of
such systems suggest that the velocity dispersion {\it within} the
dense clumps is subsonic, but that the velocity dispersion of the
clumps within the diffuse substrate is supersonic with respect to their
internal sound speed \citep[although subsonic with respect to the warm
gas;][]{KI02, Heitsch+05}.

\subsection{The Molecular Gas} \label{sec:molec_turb} 

\subsubsection{Molecular Clouds: Supersonically Turbulent, or
Collapsing?} \label{sec:non_mag_molec}

\paragraph{The evidence}

Molecular clouds (MCs) have long been known to be strongly
self-gravitating \citep[e.g.,][]{GK74, Larson81}. In view of this,
\citet{GK74} initially proposed that MCs should be in a state of
gravitational collapse, and that the observed motions in MCs (as derived
by the non-thermal linewidths of molecular lines) corresponded to this
collapse.  However, shortly thereafter, \citet{ZP74} argued against this
possibility by noting that, if all the molecular gas in the Galaxy, with
mean density $n \sim 100 \pcc$ and total mass $\Mmol \sim 10^9
\Msun$, were in free-fall, then a simple estimate of the total SF rate
(SFR) in the Galaxy, given by SFR $\sim \Mmol / \tff \sim 200~\Msun$
yr$^{-1}$, where $\tff = \sqrt{3\pi/32 G \rho}$ is the free-fall time,
would exceed the observed rate of $\sim 2~\Msun$ yr$^{-1}$
\citep[e.g.,][]{ChP11} by about two orders of magnitude. Moreover,
\citet{ZE74} argued that, if clouds were undergoing large-scale radial
motions (a regime which they assumed would include the case of a global
gravitational contraction), then the star formation activity, and the
{\sc Hii} regions associated with it, would tend to be concentrated at
the center of the cloud. Under these conditions, the H$_2$CO absorption
lines seen on the spectra of the {\sc Hii} regions, produced by the
surrounding, infalling gas, should be redshifted with respect to the CO
lines produced by the cloud as a whole, an effect which \citet{ZE74}
showed does not occur. They also argued that such a ``radial-motion''
flow regime is inconsistent with the fact that clouds contain multiple
{\sc Hii} regions, clusters, and dense clumps. A related notion, which
still persists today, is that, if a cloud is undergoing global collapse,
the largest linewidths should occur near the collapse center, as the
infall velocities should be at a maximum there, contrary to the
observation that the largest velocity dispersions occur at the largest
scales \citep{Larson81}.

These objections prompted the suggestion \citep{ZE74} that the
non-thermal motions in MCs corresponded instead to {\it small-scale} (in
comparison to the sizes of the clouds) random turbulent motions. The
need for these motions to be confined to small scales arose from the
need to solve the absence of a systematic shift between the H$_2$CO
absorption lines of {\sc Hii} regions and the CO lines from their parent
molecular clouds noted by those authors. But such a small-scale nature
also had the advantage that the turbulent ({\it ram}) pressure could
provide an approximately isotropic pressure that could counteract the
self-gravity of the clouds at large, thus providing a suitable mechanism
for keeping the clouds from collapsing and maintaining them in near
virial equilibrium \citep{Larson81}. On the other hand, because
turbulence is known to be a dissipative phenomenon \citep[e.g.,
][]{LL59}, research then focused on finding suitable sources for driving
the turbulence and avoiding rapid dissipation. The main driving source
was considered to be energy injection from stars \citep[e.g.,][see also
the reviews by Mac Low \& Klessen 2004 and V\'azquez-Semadeni
2010]{NS80, McKee89, LN06, NL07, KMM06, Carroll+09, Carroll+10,
Wang+10}, and reduction of dissipation was proposed to be accomplished
by having the turbulence being MHD, and consisting mostly of Alfv\'en
waves, which were thought not to dissipate as rapidly
\citep[e.g.,][]{SAL87}, and which could provide an isotropic pressure
\citep{MZ95}.

\begin{figure}
\begin{center}
\includegraphics[scale=.45]{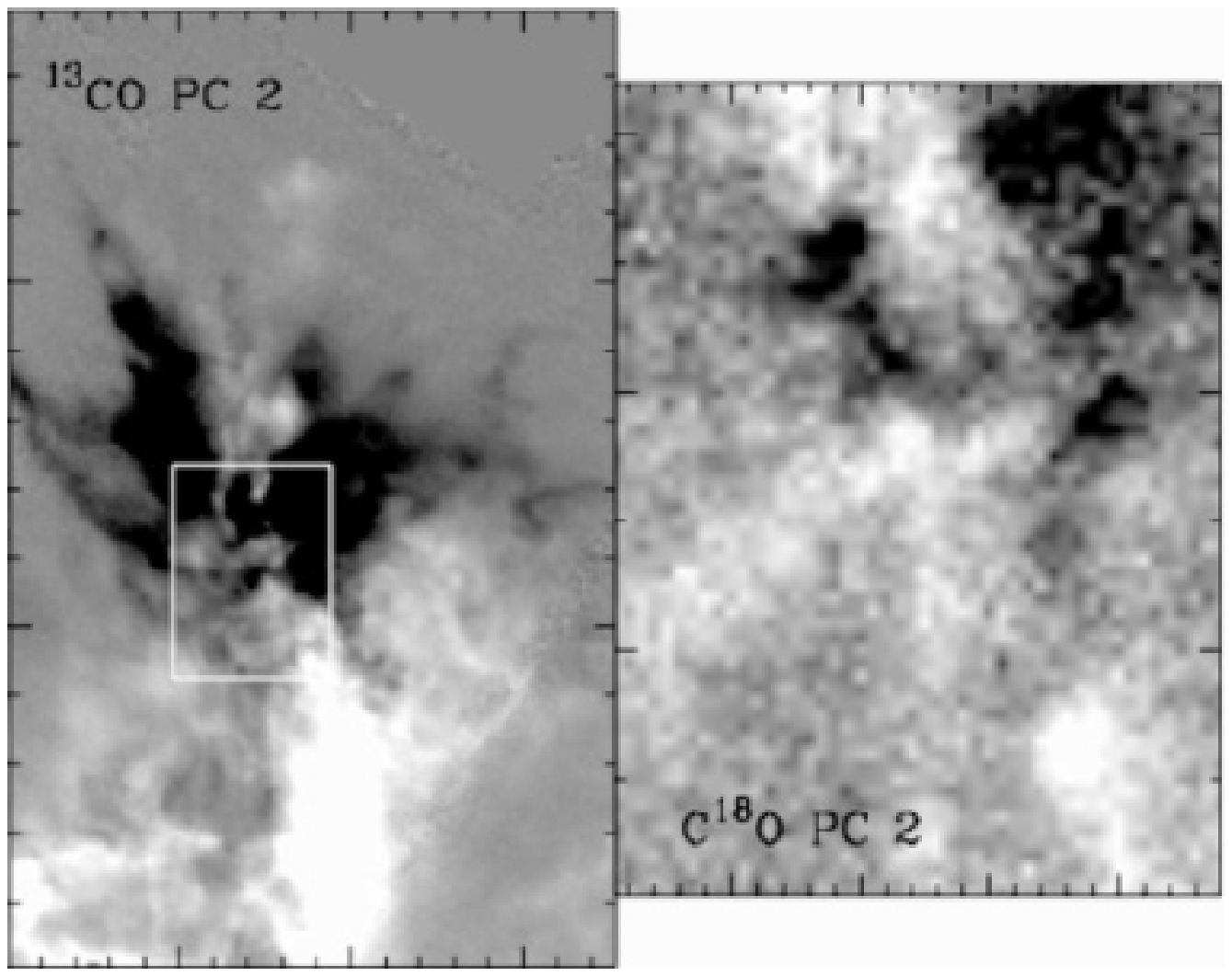}
\includegraphics[scale=.2]{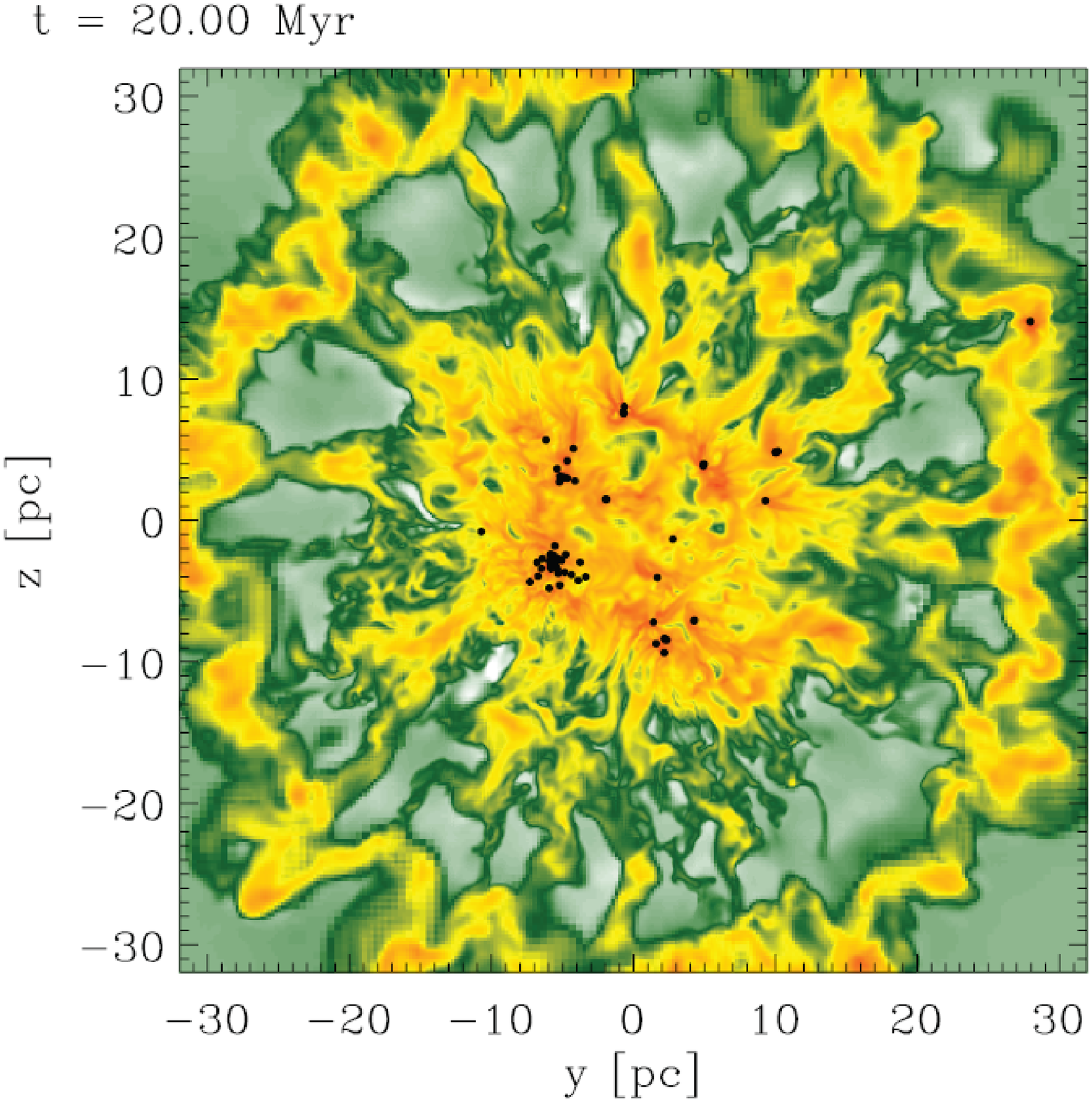}
\end{center} 
\caption{{\it Left and middle panels:} Second eigenimages obtained by
Principal Component Analysis of spectroscopic data of the star-forming
region NGC 1333, showing the main contribution to the linewidth of
molecular emission in this region \citep{Brunt+09}. The middle image
shows the region enclosed in the rectangle in the {\it left}
image. Black and white colors represent oppositely-signed components of
the velocity. \citet{Brunt+09} describe the pattern as a ``dipole'', in
which large-scale patches of alternating velocity direction are
observed. This is seen in both the large-scale ({\it left}) and the
small-scale ({\it middle}) images. {\it Right panel:} Image of the
projected density field of a 3D numerical simulation with cooling,
self-gravity, and magnetic fields, representing the formation of a dense
atomic cloud by the collision of WNM streams in the direction
perpendicular to the plane of the figure. The time shown is 20 Myr after
the start of the simulation. The black dots denote ``sink'' particles,
which replace local collapsing zones in the simulation. The whole cloud
is also collapsing, although its collapse is not completed yet by the
end of the simulation, at $t=31$ Myr. From \citet{VS+11}.}
\label{fig:dense_turb}
\end{figure}

However, in the last decade several results have challenged the
turbulent pressure-support scenario: 1) Turbulence is known to be
characterized by having the largest velocity differences occurring at
the largest scales, and MCs are no exception, exhibiting scaling
relations between velocity dispersion and size which suggest that the
largest velocity differences occur at the largest scales \citep[][Fig.\
\ref{fig:dense_turb}, {\it left} and {\it middle} panels] {Larson81, HB04,
Brunt+09}. This is inconsistent with the small-scale requirement for
turbulent support. 2) It was shown by several groups that MHD turbulence
dissipates just as rapidly as hydrodynamic turbulence \citep{ML+98,
Stone+98, PN99}, dismissing the notion of reduced dissipation in
``Alfv\'en-wave turbulence'', and thus making the presence of strong
driving sources for the turbulence an absolute necessity. 3) Clouds with
very different contributions from various turbulence-driving mechanisms,
including those with little or no SF activity, such as the
so-called {\it Maddalena's cloud} \citep{MT85}, show similar turbulence
characteristics
\citep{Williams+94, Schneider+11}, suggesting that stellar energy
injection may not be the main source of turbulence in MCs.

Moreover, simulations of dense cloud formation in the nonmagnetic case
have shown that, once a large cold CNM cloud forms out of a collision of
WNM streams, it quickly acquires a large enough mass that it can begin
to collapse gravitationally in spite of it being turbulent \citep{VS+07,
VS+10, HH08, Heitsch+08a}. This happens because, as the atomic gas
transitions from the warm to the cold phase, its density increases by
roughly two orders of magnitude, while the temperature drops by the same
factor. Thus, the Jeans mass in the gas, proportional to the product
$n^{-1/2} T^{3/2}$, drops by a factor $\sim 10^4$, implying that the
cold cloud assembled by the compression can rapidly exceed its Jeans
mass. 

In turn, the gravitational contraction very effectively enhances the column
density of the gas, promoting the formation of molecular hydrogen (H$_2$)
\citep{HBB01, Bergin+04, HH08}, and so it appears that the formation of a
{\it molecular} cloud may involve some previous gravitational contraction
\citep[see also][]{McKee89}. In addition, according to the discussion in
Secs.\ \ref{sec:cloud_form} and \ref{sec:atomic_turb}, the CNM clouds
formed by converging WNM flows should be born turbulent and clumpy. This
turbulent nature of the clouds further promotes the formation of
molecular hydrogen \citep{GM07a}. The simulations \citep{VS+07, VS+10,
HH08} show that the nonlinear, turbulent density fluctuations can
locally complete their collapse {\it before} the global collapse of the
cloud is completed (Fig.\ \ref{fig:dense_turb}, {\it right} panel), both
because their densities are large enough that their free-fall time is
significantly shorter than that of the whole cloud \citep{HH08, Pon+11},
and because clouds in general have flattened or filamentary shapes
\citep[e.g.,][] {Bally+89, deGeus+90, HT03, Molinari+10,
Andre+10}. Interestingly, the free-fall time for these geometries may be
much larger than the standard free-fall time, $\tff = \sqrt{3\pi/32 G
\rho}$, which is applicable to a spherically symmetric structure
\citep{Toala+12, Pon+12}. Thus, an approximately spherical clump of the
same volume density within a flattened or elongated structure can
collapse much earlier than the non-spherical cloud in which it is
immersed.

In addition, the turbulent velocities initially induced in the clouds by
the converging flows in the simulations are observed to be relatively
small (only moderately supersonic [$\Ms \sim 3$] with respect to the
dense gas). Strongly supersonic ($\Ms \sim 10$) velocities like those
observed in real molecular clouds only develop later in the simulations,
due to the ensuing gravitational contraction \citep{VS+07}. This is in
agreement with the fact that CNM
clouds are observed to typically have moderately supersonic ($\Ms \sim
3$) velocity dispersions \citep{HT03}, while GMCs are observed to have
much larger turbulent rms Mach numbers, $\Ms \sim 10$--20
\citep{Wilson_etal70}.  Finally, \citet{Banerjee+09} noted that, in
their numerical simulations, the clumps with highest internal velocity
dispersions were those that had already formed collapsed objects
(``sink'' particles), even though energy feedback from the sinks was not
included. This again suggested that the largest velocities develop by
the action of self-gravity.

\paragraph{Do observations rule out global gravitational contraction in
star-forming molecular clouds?}

It is very important to note that the possibility of MCs being
in gravitational collapse is not in contradiction with any observed
properties of MCs. First, as noted by \citet{BP+11a}, the magnitudes of
the virial and free-fall velocities for a self-gravitating object are
observationally indistinguishable. Thus, the interpretation of cloud
energetics in terms of virial equilibrium is completely interchangeable
by an interpretation of collapse. 

Also, one important argument against the
possibility of gravitational collapse of MCs is the argument by
\citet{ZP74} that it would lead to exceedingly large SFRs. We discuss
the possible resolution of this conundrum in \S \ref{sec:feedbk}.
Another frequent argument against the global gravitational contraction
scenario is that such a regime should produce readily observable
signatures, such as the systematic shift between the CO lines from the
clouds and the absorption lines seen towards {\sc Hii} regions proposed
by \citet{ZE74}, as discussed at the beginning of this section.
However, it should be noted that the argument by ZE74 against
large-scale motions in the clouds would also apply to turbulent motions
as we presently understand them, since turbulent flows in general have
the largest velocity differences across the largest velocity
separations, as discussed above in relation to the left and middle
panels of Fig.\ \ref{fig:dense_turb}. The only kind of turbulence that
would not be invalidated by ZE74's argument would be microscopic
turbulence, in which the largest turbulent scale should be much smaller
than the size of the cloud, but, as already discussed above, this is
clearly not the case in molecular clouds, as illustrated by the {\it left} and
{\it middle} panels of Fig.\ \ref{fig:dense_turb}.

Moreover, the above arguments against global collapse in clouds are
based on the assumptions that the cloud has a roughly spherical
symmetry, and that the collapse is monolithic, meaning that there is a
single, dominant flow, aimed at a major, localized collapse
center. Actually, numerical simulations of cloud formation and evolution
show that this is not the case. As mentioned above, the clouds are far
from having a spherical symmetry, and instead tend to have flattened or
filamentary shapes. In addition, the clouds are
born turbulent, and therefore they contain nonlinear density
fluctuations, which have shorter free-fall times than the average in the
cloud, and thus collapse earlier. Thus, multiple collapse centers arise
in the cloud before the global collapse is completed, and, as a
consequence, there is no single, evident, dominant collapse center,
possibly resolving the concerns of
\citet{ZE74}.  Essentially, the cloud fragments gravitationally, with
the local collapse centers accreting from filaments that in turn accrete
from the bulk of the cloud (G\'omez
\& V\'azquez-Semadeni, in prep.), in agreement with the structure of
clumps and their surrounding filaments \citep{Myers09, Andre+10,
Palmeirim+13}. Towards the end of the evolution, the locally collapsing
regions formed earlier tend to merge to form a massive region then
acquires large densities and velocities, typical of massive-star forming
regions \citep{VS+09}. This flow regime has been termed {\it
hierarchical, chaotic gravitational fragmentation}
\citep{VS+09, BP+11a}.

Note, however, that all of the above arguments in favor of gravitational
contraction motions in MCs probably apply mostly to clouds in the
early-to-intermediate stages of their evolution; that is, from their
formation to their strongly star-forming stages. Nevertheless, after
strong stellar feedback has disrupted the clouds, it is likely that
shreds may remain that may remain in a relatively quiescent stage,
perhaps supported by the magnetic field, without forming stars, and
perhaps even being on their way to dispersal \citep{Elm07, VS+11}.

All of the above evidence suggests that the observed supersonic
nonthermal motions in MCs may evolve from being dominated by random
turbulence in the early evolutionary stages of the (mostly atomic)
clouds, to being infall-dominated at more advanced (mostly molecular) stages,
characterized by large densities, velocities, and star formation rates.
Note, however, that the turbulent component may be maintained or even
somewhat amplified by the collapse \citep{VS+98, RG12}.

In this scenario of hierarchical gravitational fragmentation, the main
role of the truly turbulent (i.e., fully random) motions is to provide
the nonlinear density fluctuation seeds that will collapse locally once
the global contraction has caused their density to increase sufficiently
for them to become locally gravitationally unstable
\citep{CB05}. Evidence for such multi-scale collapse has recently begun
to be observationally detected \citep{Galvan+09, Schneider+10}.

\subsubsection{The Molecular Gas. Results Including the Magnetic Field}
\label{sec:molec_turb_magn} 

According to the discussion in \S \ref{sec:cloud_form_mag}, the
formation of a cold, dense {\it atomic} cloud can be accomplished by the
compression of warm material along magnetic field lines, which nonlinearly
triggers a phase transition to the cold phase. However, as discussed
in the previous section, the formation of a {\it molecular} cloud
probably requires the gravitational contraction of the atomic cloud
previously formed by the compression. Thus, in the presence of the
magnetic field, this requires an understanding of the role of magnetic
support; that is, of the evolution of the mass-to-flux ratio (MFR).

As is well known, and was reviewed in \S \ref{sec:MFR_evol}, there
exists a critical value of the MFR below which the magnetic field is
able to support the cloud against its own self-gravity. 
Along the direction of the field
lines, the criticality condition in terms of the mass column density
$\Sigma = \rho L$ and the field strength $B_0$ for a cylindrical
geometry is \citep{NN78},
\beq
\left(\Sigma/B_0\right)_{\rm crit} = (4 \pi^2 G)^{-1/2}  \approx
0.16~ G^{-1/2}, 
\label{eq:NN78}
\eeq
where $\rho$ is the mass density and $L$ is the cylinder length.  This
condition gives the accumulation length, in terms of fiducial values
representative of the ISM in the solar neighborhood, as \citep{HBB01}
\begin{equation}
\Lc \approx 470 \left(\frac{B_0}{5 \lambda{\rm G}} \right) \left(\frac{n}{1
\pcc} \right)^{-1}~{\rm pc},
\label{eq:acc_length}
\end{equation}
where we have assumed $\mum = 1.27$. In principle, if the Galactic field
is primarily azimuthal, then the Galactic ISM at large is magnetically
supercritical in general, because field lines circle around the entire
Galactic disk, and thus sufficiently long distances are always available
along them.\footnote{Note, however, that supercriticality does not
necessarily imply collapse, since the gas may be thermally or otherwise
supported, as is likely the case for the diffuse warm medium at scales
of hundreds of parsecs.} Thus, {\it the MFR of a system is not a
uniquely defined, absolute parameter, but rather depends on where the
boundaries of the system are drawn.} Also, recall that the critical value of
the MFR depends on the local geometry of the system being considered.
For instance, a system with spherical symmetry has a critical value of 
$(\Sigma/B_0)_{\rm crit} = (6\pi^2 G)^{-1/2} \approx 0.13 ~ G^{-1/2}$
\citep[e.g.][]{Shu92}, somewhat smaller than that given by eq.\
(\ref{eq:NN78}). 

Now consider a cloud or clump that is formed by the accumulation of gas
along field lines in general.\footnote{Since compressions perpendicular
to the magnetic field cannot induce collapse of an initially subcritical
region, as they do not change the MFR, and compressions oblique to the
field can produce collapse by reorienting the directions of the flow and
the field lines \citep{HP00}, our assumed configuration involves no loss
of generality.} In the rest of this discussion, we will generically
refer to the resulting density enhancement as a ``cloud'', referring to
either a cloud, a clump, or a core. Although redistribution of matter
along field lines does not in principle affect the {\it total} MFR along
the full ``length'' of a flux tube, this length is a rather meaningless
notion, since the flux tube may extend out to arbitrarily long
distances. What is more meaningful is the MFR {\it of the dense gas that
makes up the cloud}, since the cloud is denser than its surroundings,
and thus it is the main source of the self-gravity that the field has to
oppose. In fact, for the formation of a cloud out of flow collisions in
the WNM, the density of the cloud is $\sim 100$ times larger than that of the
WNM \citep{FGH69, Wolfire+95}, and so the self-gravity of the latter is
negligible. Thus, in this problem, natural boundaries for the cloud are
provided by the locus of the phase transition front between the dense
and the diffuse gas, allowing a clear working definition of the MFR.

However, contrary to the very common assumption of a constant cloud
mass, the formation of clouds by converging gas streams implies that the
mass of the cloud is a (generally increasing) function of time
(cf.\ \S \ref{sec:clump_nature}), a conclusion 
that has recently been reached observationally as well
\citep{Fukui+09}. This means that, {\it within the volume of 
the cloud, the MFR is also an increasing quantity}, since the flux
remains constant if the flow is along field lines, while the mass
increases \citep[see also][]{Shu+07}. If the cloud starts from
essentially zero mass, this in turn implies that the MFR of a cloud is
expected to start out strongly subcritical (when the cloud is only
beginning to appear), and to evolve towards larger values at later
times. Rewriting eq.\ (\ref{eq:acc_length}) for the column density, we
see that the cloud becomes supercritical when \citep{VS+11}
\begin{equation}
N_{\rm cr} \approx 1.5 \times 10^{21} \left(\frac{B_0}{5 \mu {\rm G}}
\right) \psc,
\label{eq:Sigma_crit}
\end{equation}
where $N \equiv \Sigma/\mum m_H$ is the {\it number} column density,
and is to be measured along the field lines. 
The critical column density for magnetic criticality given by eq.\
(\ref{eq:Sigma_crit}) turns out to be very similar, at least for solar
neighbourhood conditions, to the critical column density of hydrogen
atoms necessary for cold atomic gas to become molecular, $N_{\rm H}
\sim 1$--2$ \times 10^{21}\psc$ \citep[e.g.,][]{FC86, vDB88, vDB98,
HBB01, GM07a, GM07b, Glover_etal10}.

Moreover, the critical column density given by eq.\
(\ref{eq:Sigma_crit}) is also very similar to that required for
rendering cold gas gravitationally unstable, which is estimated to be
\beq
N_{\rm grav} \approx 0.7 \times 10^{21} \left( \frac{P/k}
{3000~ K \pcc} \right)^{1/2}\psc
\label{eq:N_self-grav}
\eeq
\citep{FC86, HBB01}.  Thus, {\it the evolution of a cloud
is such that it starts out as an atomic, unbound, and subcritical
diffuse cloud \citep{VS+06} and, as it continues to accrete mass from
the warm atomic medium, it later becomes molecular, supercritical, and
collapsing, at roughly the same time} \citep{HBB01}. 
This is fully consistent with
the observation that diffuse atomic clouds are in general strongly
subcritical \citep{HT05} and not strongly self-gravitating, while GMCs
are approximately critical or moderately supercritical
\citep{Crutcher99, Bourke+01, TC08}, and are generally
gravitationally bound \citep[e.g., ][]{Blitz93}.

It is important to note that this is in stark contrast to the SMSF
\citep[see, e.g., the reviews by][]{SAL87, Mousch91}, where it was
considered that the magnetic criticality of a cloud was the main
parameter determining whether it would form only low-mass stars and at a
slow pace (in the case of subcritical clouds), or form clusters,
including high-mass stars, and at a fast pace (supercritical
clouds). This constituted a bimodal scenario of SF, and sub- and
supercritical clouds constituted two separate classes.

Instead, in the evolutionary scenario for MCs described above, clouds
are expected to {\it evolve} from being simultaneously atomic,
subcritical and not strongly self-gravitating to being molecular,
supercritical and strongly self-gravitating. Next, the roughly
simultaneous transition to self-gravitating and supercritical suggests
that, in general, GMCs
should be in a state of gravitational contraction,
at least initially, even in the presence of typical magnetic field
strengths in the Galactic disk. Of course, significant scatter in the
MFR is expected, both intrinsically (see \S \ref{sec:B-n_corr}) and as a
consequence of observational uncertainties \citep[e.g.,][]{Crutcher99},
and thus a certain fraction of the GMCs may remain subcritical up to
significantly evolved stages, or even throughout their entire
evolution. This case is discussed further below.


The formation and evolution of molecular clouds in the magnetic case has
been investigated recently using numerical simulations of GMC formation
by compressions in the WNM aligned with the magnetic field
\citep{Hennebelle+08, Banerjee+09, VS+11}. The latter authors in
particular included self-gravity and AD, and considered three cases: one
supercritical, with $\lambda =1.3$, and two subcritical, with $\lambda =
0.9$ and 0,7, corresponding to mean field strengths of 2,
3, and 4 $\mu$G, respectively. The initial magnetic field was considered
uniform. In all cases, the mean density was $1 \pcc$ and the temperature
$T= 5000$ K. The compressions consisted of two oppositely-directed
streams of gas at the mean density, and of length 112 pc, immersed in a
256-pc box.

The evolution of the subcritical cases is worth discussing in detail, as
it differs somewhat from simple expectations. These simulations produced
a dense cloud that quickly began to contract gravitationally, similarly
to non-magnetic simulations. This occurred because a uniform magnetic
field does not provide any support, since the latter requires the
existence of a magnetic gradient. Support builds up gradually as the
field lines are bent. The clouds thus contracted for a few tens of Myr,
until the magnetic tension was large enough to halt the collapse, at
which point they re-expanded, and entered an oscillatory regime, around
the equilibrium configuration. However, due to the existence of
diffusion (both numerical and from AD), local collapse events managed to
occur, in agreement with the notions from the SMSF \citep[see
also][]{McKee89}. The notable difference with that model, though,
occurred in the fact that the clouds only formed stars during the global
contraction phase, especially at maximum compression, and essentially
shut off in the re-expanding phase. This is in contrast to the SMSF, in
which the GMCs at large were assumed to be in equilibrium and forming
stars continuously, albeit slowly.

The above discussion suggests that the possibility of {\it star-forming}
molecular clouds being in a state of gravitational contraction may hold
even if they have subcritical MFRs. The subsequent re-expansion of
these clouds (or their remnants) may lead to a star-formation-inactive
and quiescent phase, perhaps on their way to dispersal, if the clouds
are exiting the spiral arms by that time, as proposed by \citet{Elm07}.

\section{Star Formation in the Turbulent ISM} \label{sec:turb_SF}

\subsection{Does Molecular Cloud ``Turbulence'' Provide Support for
Molecular Clouds?} \label{sec:turb_grav?} 

In the previous sections we have discussed how large-scale compressions
in the general ISM produce density fluctuations, in particular by
nonlinearly inducing phase transitions from the cold to the warm
medium. Because the largest dimensions of the clouds thus formed are as
large as the transverse dimension of the compression that formed them,
and because of the large drop in the local Jeans mass upon the phase
transition (cf.\ Sec.\ \ref{sec:non_mag_molec}),
they can soon find themselves being strongly gravitationally unstable
and proceed to collapse. It is important to note that the
large-scale compression forming the clouds may (and in fact, is likely
to) have an origin different from the general turbulence in the ISM,
such as, for example, large-scale instabilities like the magneto-Jeans
one \citep[e.g.][]{KO01}, or simply the passage of the stellar spiral-arm
potential well. 

In fact, it is worth noting that just the turbulence driven by
supernovae does not seem to be able to sustain itself, since the mass
driven into a Jeans-unstable regime per unit time by the turbulence is
not enough to maintain the same supernova rate that drives the
turbulence \citep{JM06}. This conclusion is also supported by the fact
that successive generations of triggered SF do not appear to
be able to form stars as massive as in the previous generation
\citep{DZ11}. Thus, it appears safe to conclude that the main driver of star
formation is gravity at the largest scales.

During the last decade, the main role of interstellar turbulence has
been thought to be the {\it regulation} of SF, mainly on the basis of
the assumption that the turbulent velocity dispersion contributes to the
support of molecular clouds against their self-gravity, analogously to
the role of the thermal velocity dispersion, and perhaps including a
scale-dependent amplitude \citep{Chandra51, Bonazzola+87, BM92, VG95,
MK04, KM05, HC08, HC11, PN11}. Thus, turbulence has been thought to provide
support to clouds as a whole, while simultaneously inducing small-scale
density fluctuations (clumps) within the clouds that may undergo
gravitational collapse if they are compressed enough for their Jeans
mass to become smaller than their actual mass \citep{VBK03, BP+07}. In
this manner, global collapse of the clouds could be prevented, avoiding
the \citet{ZP74} conundrum that the global collapse of molecular clouds
would cause an excessively large SFR (see \S \ref{sec:non_mag_molec}),
while at the same time allowing for the collapse of a small fraction of
the mass, brought to instability by the local supersonic turbulent
compressions. However, this last notion was challenged by \citet{CB05},
who argued that the turbulence only provides the seed density
fluctuations for subsequent gravitational fragmentation, without
significant local reductions in the Jeans mass induced by the
turbulence.

\begin{figure}
\begin{center}
\includegraphics[scale=.3]{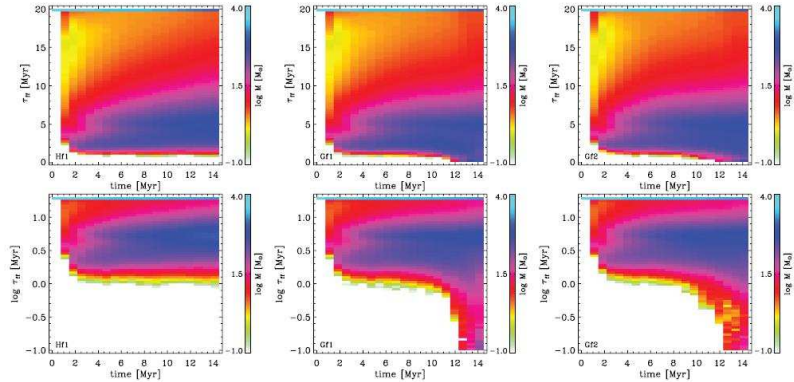}
\end{center}
\caption{Time evolution of the mass (indicated by the color scale) at a
given free-fall time (vertical axis) in numerical simulations of cloud
formation and evolution by \citet{HH08}. The {\it top} row shows the
free-fall time in a linear scale, while the {\it bottom} row shows it in
logarithmic scale. The panels on the {\it left} show a simulation with
no self-gravity, while the {\it middle} and {\it right} panels show two
different simulations with self-gravity. In the case with no
self-gravity, the fraction of mass at a given free-fall time is seen to
remain nearly constant, and the minimum free-fall time to remain at
$\sim 1$ Myr, while in the cases with self-gravity, the minimum
free-fall time decreases secularly. Note that, in these plots, $\tff$
is simply a proxy for the density, since $\tff \propto \rho^{-1/2}$, and
so it can be evaluated even if gravity is not included in the simulations.}
\label{fig:HH08_f5}
\end{figure}

Moreover, \citet{HH08} showed that the fraction
of mass with short free-fall times ($\lesssim 1$ Myr) in the clouds
increases monotonically over time in the presence of self-gravity,
indicating a secular evolution towards higher densities, while
simulations with no self-gravity exhibited a stationary
fraction of mass with short free-fall times, as would be the case
in clouds supported against collapse by the turbulence (Fig.\
\ref{fig:HH08_f5}).

Finally, the simulations have also shown that the fraction of molecular
gas also increases in time, so that the cloud would indeed be classified
as atomic in its early phases, and as molecular in later ones
\citep{HH08, Clark+12}. In particular, the latter authors have shown
that the formation of CO-dominated regions only occurs $\sim 2$ Myr
before SF starts, although significant amounts of H$_2$ can appear
earlier.

All of the above evidence suggests that the strongly supersonic motions
observed in MCs may be a {\it manifestation} of the gravitational
contraction occurring in the clouds, rather than truly turbulent (i.e.,
random) motions, of a separate origin, that can
counteract the gravitational contraction of the clouds. It is worth noting
here that \citet{KH10} have recently shown that, in general, the
accretion power at scales from entire Galactic disks to protostellar
disks, passing through the GMC scale, is more than enough to drive the
turbulence observed in these systems. However, it should be noted that,
in the case of GMCs, this suggestion differs qualitatively from the
nature of the motions discussed above. Rather than accretion {\it
driving} turbulent motions in the clouds which can then support them,
the discussion above suggests that the observed motions in the clouds
{\it are} the infall itself, with only a small, subdominant, truly
random turbulent component superposed on them. In this case, these
motions cannot provide support against the self-gravity of the clouds.

\subsection{Regulation of Star Formation Via Stellar Feedback}
\label{sec:feedbk}

All of the above evidence strongly suggests that interstellar clouds
undergo a secular evolution, starting their existence as moderatly
supersonic, magnetically subcritical, sheet-like atomic clouds, and
evolving towards becoming supercritical, molecular, gravitationally
contracting objects.  However, in this case, the \citet{ZP74} SF
conundrum (cf.\ \S \ref{sec:non_mag_molec}) must be addressed. That is,
if MCs are essentially in free-fall, how to prevent the SFR from being
two orders of magnitude larger than it is observed to be in the Galaxy?

Early studies proposed that ionizing radiation from massive stars
should be able to disperse a cloud as early as when only $\sim
10$\% of the cloud's mass has been converted to stars
\citep[see, e.g., Sec.\ 4 of][]{Field70}, so that the remaining 90\%
would be prevented from forming any more stars. This suggestion,
however, was challenged by \citet{Mousch76}, who argued that those
estimates were based on the assumption of unrealistically low mean
densities for the clouds ($\sim 10 \pcc$), and that using more realistic
values ($\sim 10^5 \pcc$) would result in a grossly insufficient amount
of ionization in the cloud, thus invalidating the mechanism as a
suitable one for dispersing the clouds. As an alternative,
\citet{Mousch76} proposed the basic notions for the SMSF: that the
MCs should be magnetically subcritical in general, so that their
envelopes would remain supported by the magnetic tension, while only the
central core would be able to proceed to collapse through AD (cf.\ \S\S
\ref{sec:eqs} and \ref{sec:B-n_corr}). However, observational evidence
from the last decade has suggested that most MCs are likely to be at least
moderately magnetically supercritical \citep[e.g.,][]{Bourke+01, TC08,
Crutcher+10}, a conclusion also reached by theoretical arguments (see \S
\ref{sec:molec_turb_magn} and references therein).

Another alternative was the proposal that MCs could be supported by
turbulence, either hydrodynamical or MHD. However, since turbulence
needs to be continuously driven, two variants have been considered for
the driving: either it might be due to feedback from stellar sources
internal to the clouds (cf.\ \S \ref{sec:non_mag_molec}), or else to
external driving sources such as supernova shocks. However, as discussed
in \S\S \ref{sec:cloud_form_hd} and \ref{sec:molec_turb_magn}, the role
of external turbulence seems more likely to be the driving of MC {\it
formation}, rather than the driving of the strongly supersonic internal
turbulence of the GMCs, because the turbulence induced in the
forming clouds is only moderately supersonic, rather than strongly so
(cf.\ \S \ref{sec:turb_grav?}).

The possibility of driving the turbulence by stellar feedback from
inside the clouds has been extensively studied, both analytically and
numerically \citep[e.g.,][]{NS80, McKee89, LN06, NL07, Carroll+09,
Carroll+10, Wang+10}. In most such studies, it has been concluded that
this feedback can maintain the clumps within GMCs in near virial
equilibrum. Studies of the SFR and the SFE under these conditions have
often idealized the turbulence as being simply randomly driven, and have
shown that in this case the SFE can be maintained at levels of a few
percent, comparable to the observed ones \citep[e.g.,][]{KHM00, VS+03,
VS+05b}.

However, as discussed in the review by \citet{VS10}, numerical
simulations of the momentum feedback from protostellar outflows have
only considered numerical boxes at the parsec (clump) scale, neglecting
the infall from the environment of the clump, which has been observed in GMC
formation simulations. This adds a large amount of ram pressure to the
system not included in those simulations.  Thus, it seems that outflows
cannot provide sufficient feedback to prevent the collapse of entire
GMCs.

The role of massive-star ionization feedback in the support of GMCs has
been investigated semi-analytically by \citet{KMM06} and
\citet{Goldbaum+11} considering the time-dependent virial theorem in the
presence of feedback, and of feedback and infall, respectively,
concluding that the clouds may oscillate around the virial equilibrium
state for several Myr, until they are finally dispersed. However,
full numerical simulations of this problem \citep{VS+10} suggest
that the infall is not suppressed, and instead that the regulation of
the SFR occurs because most of the infalling material is evaporated
before it can form further stars, except in the case of the most massive
($\sim 10^6 \Msun$) GMCs \citep{Dale+12}, where supernova feedback may
be also required to accomplish the dispersal of the clouds.

Thus, it appears that the resolution of the \citet{ZP74} conundrum lies
not in the prevention of the global contraction of star-forming GMCs,
but rather on the effect of the feedback, and that this effect is, after
all, essentially as initially suggested by \citet{Field70}. The
resolution of the objection by \citet{Mousch76}, in turn, appears to lie
in that the fraction of mass that is at very high densities ($> 10^5
\pcc$) is very small \citep[see, e.g.,][secs.\ VII and IX]{Blitz93} and
inhomogeneously distributed, so that eventually {\sc Hii} regions may
break out from the densest regions and ionize the rest of the MC
\citep[see, e.g.,][]{Peters+10}.

\section{Summary and Conclusions} \label{sec:conclusions}

In this contribution, we have briefly reviewed the role and interaction
between the main physical processes present in the ISM: radiative
heating and cooling, magnetic fields, self-gravity, and turbulence, and
their implications for the SF process. The
presence of radiative heating and cooling implies in general that the
gas behaves in a non-isentropic (i.e., non-adiabatic) way, and in
particular it may become {\it thermally unstable} in certain regimes of
density and temperature, where low-amplitude (i.e., {\it linear}) perturbations
can cause runaway heating or cooling of the gas that only stops when the
gas exits that particular regime. This in turn causes the gas to avoid
those unstable density and temperature ranges, and to settle in the
stable ones, thus tending to segregate the gas into different phases of
different densities and/or temperatures. In classical models of the ISM,
only the stable phases were expected to exist in significant amounts.

We then discussed some compressible MHD turbulence basics, and the
production, nature and evolution of turbulent density fluctuations in
polytropic (i.e., of the form $P \propto \rho^\gamef$) flows, discussing
in particular the probability density function (PDF) of the density
fluctuations, which takes a lognormal form in isothermal regimes, and
develops power-law tails in polytropic ones. We also discussed the
correlation (and, at low densities, lack thereof) between the magnetic
field and the density as a consequence of the superposition of the
different MHD wave modes, and the evolution of the mass-to-magnetic flux
ratio (MFR) as density enhancements are assembled by turbulent
fluctuations. 

We next discussed turbulence in the multi-phase ISM, noting that, since
turbulence is an inherently mixing phenomenon, it opposes the
segregating effect of thermal instability, causing the production of gas
parcels in the classically forbidden unstable regimes, which may add up
to nearly half the mass of the ISM, although the density PDF in general
still exhibits some multimodality due to the preference of the gas to settle
in the stable regimes. The existence of gas in the unstable ranges has
been established by various observational studies.

Next, we discussed the nature of the turbulence in the different ranges
of density and temperature of the gas, noting that in the diffuse
ionized regions, where the flow is transonic (i.e., with Mach numbers
$\Ms \sim 1$), the gas appears to behave in an essentially
incompressible way, exhibiting Kolmogorov scalings over many orders of
magnitude in length scale. However, in the neutral atomic component,
where the gas is thermally bistable, the flow is expected to exhibit
large density and temperature fluctuations, by up to factors $\sim 100$,
thus being highly fragmented. We also pointed out that large-scale
compressions in the warm neutral gas, which may be triggered by either
random turbulent motions, or by yet larger-scale instabilities, may
nonlinearly induce the formation of large regions of dense, cold gas;
much larger, in particular, than the most unstable scales of TI, which
have sizes $\sim 0.1$ pc, thus forming large cold atomic clouds that may
be the precursors of giant molecular clouds (GMCs). This is because
these clouds are expected to become molecular, gravitationally unstable,
and magnetically supercritical at approximately the same time, so that
when they reach a mostly molecular stage, they are likely to be
undergoing generalized gravitational contraction. 

The clouds are born internally turbulent and clumpy, and the resulting
nonlinear density fluctuations (``clumps'') eventually become
locally gravitationally unstable during the contraction of the whole
large-scale cloud. Because they are denser, they have shorter free-fall
times, and can complete their local collapses before the global one
does, thus producing a regime of {\it hierarchical gravitational
fragmentation}, with small-scale, short-timescale collapses occurring
within larger-scale, longer-timescale ones. It is thus quite likely that
the flow regime in the dense molecular clouds corresponds to a dominant
multi-scale gravitational contraction, with smaller-amplitude random
(turbulent) motions superposed on it. 

The local collapses cause star formation (SF) that begins before the
global collapse is concluded, and the ionizing feedback from the massive
stars that form during this stage appears to be sufficient to erode and
disperse the clouds before the entire mass of the clouds is converted to
stars, thus avoiding the objection by \citet{ZP74} to free-falling GMCs,
that they would form stars at a rate much larger than the observed
Galactic rate. They do so, but only for short periods of time, before
most of their mass gets dispersed.

We conclude that turbulence in the magnetized, multi-phase,
self-gravitating ISM is an extremely rich and complex phenomenon, but
whose (thermo)dynamics is beginning to be understood, together with its
relation to the star formation process.

\begin{acknowledgement}
This work has been funded in part by CONACYT grant 102488.
\end{acknowledgement}

\end{document}